\documentclass[journal]{IEEEtran}
\IEEEoverridecommandlockouts

\usepackage{xcolor,soul,framed} 
\usepackage{algorithmic}
\usepackage{algorithm}
\colorlet{shadecolor}{yellow}
\usepackage[pdftex]{graphicx}
\graphicspath{{../jpeg/}}
\DeclareGraphicsExtensions{.pdf,.jpeg,.png}
\usepackage{color,soul}
\usepackage[cmex10]{amsmath}
\usepackage{amssymb}
\usepackage{array}
\usepackage{mdwmath}
\usepackage{mdwtab}
\usepackage{eqparbox}
\usepackage{url}
\usepackage{bm}
\usepackage{amssymb}
\usepackage{pifont}
\usepackage{multirow}  
\usepackage{multicol}  
\usepackage {cite}
\usepackage{graphicx}
\newcommand{\Min}{\operatorname{Minimize}}
\newcommand{\Max}{\operatorname{Maximize}}

\usepackage{acro}

\DeclareAcronym{IRS}{
  short = IRS,
  long  = intelligent reﬂecting surface}
  
\DeclareAcronym{NLoS}{
  short = NLoS,
  long  = non-line-of-sight}

 \DeclareAcronym{LoS}{
  short = LoS,
  long  = line-of-sight}
  
\DeclareAcronym{DoA}{
  short = DoA,
  long  = direction of arrival}
  
\DeclareAcronym{BF}{
  short = BF,
  long  = beamforming}
  
\DeclareAcronym{BS}{
  short = BS,
  long  = base station}
  
\DeclareAcronym{ISAC}{
  short = ISAC,
  long  = integrated sensing and communication}
  
\DeclareAcronym{SC}{
  short = SC,
  long  = semantic communication}
  
\DeclareAcronym{ISASC}{
  short = ISASC,
  long  = integrated sensing and semantic communication}

\DeclareAcronym{EVE}{
  short = EVE,
  long  = eavesdropper}

\DeclareAcronym{MST}{
  short = MST,
  long  = malicious sensing target}

\DeclareAcronym{AN}{
  short = AN,
  long  = artificial noise}

\DeclareAcronym{DSS}{
  short = DSS,
  long  = dedicated sensing signal}
  
\DeclareAcronym{CRB}{
  short = CRB,
  long  = Cramér-Rao bound}
  
\DeclareAcronym{SDP}{
  short = SDP,
  long  = semi-definite programming}
  
\DeclareAcronym{GSS}{
  short = GSS,
  long  = golden-section search}

\DeclareAcronym{GRM}{
  short = GRM,
  long  = Gaussian randomization method}

\DeclareAcronym{PLS}{
  short = PLS,
  long  = physical layer security }

\DeclareAcronym{SSR}{
  short = SSR,
  long  = semantic secrecy rate}

\DeclareAcronym{SecR}{
  short = SecR,
  long  = secrecy rate}

\DeclareAcronym{SemR}{
  short = SemR,
  long  = semantic rate}
  
\DeclareAcronym{SCU}{
  short = SCU,
  long  = semantic communication user}

\DeclareAcronym{SNR}{
  short = SNR,
  long  = signal-to-noise ratio}
  
\DeclareAcronym{SINR}{
  short = SINR,
  long  = signal-to-interference-plus-noise ratio}
  
\DeclareAcronym{SS}{
  short = SS,
  long  = semantic similarity}

\DeclareAcronym{SCA}{
  short = SCA,
  long  = successive convex approximation}

\DeclareAcronym{SDR}{
  short = SDR,
  long  = semi-definite relaxation}

\DeclareAcronym{MOOP}{
  short = MOOP,
  long  = multi-objective optimization problem}

\DeclareAcronym{SOOP}{
  short = SOOP,
  long  = single-objective optimization problem}

\begin{document}
\bstctlcite{IEEEexample:BSTcontrol}
    \title{A Physical Layer Security Framework for IRS-Assisted Integrated Sensing and\\ Semantic Communication Systems}
    
    \author{Hamid~Amiriara,
            Mahtab~Mirmohseni,~\IEEEmembership{Senior Member,~IEEE},
            Ahmed~Elzanaty,~\IEEEmembership{Senior Member,~IEEE},
            Yi~Ma,~\IEEEmembership{Senior Member,~IEEE},
            and Rahim~Tafazolli,~\IEEEmembership{Fellow,~IEEE}
   \thanks{Part of this paper has been accepted at the 2025 IEEE Wireless Communications and Networking Conference, 24–27 March 2025, Milan \cite{WCNC2025}.}
    \thanks{The authors are with 5/6GIC, the Institute for Communication Systems, University of Surrey, GU2 7XH Guildford, U.K. (e-mail: \{h.amiriara, m.mirmohseni, a.elzanaty, y.ma, r.tafazolli\}@surrey.ac.uk).}
    \thanks{This work is supported by the UK Department for Science, Innovation and Technology under the Future Open Networks Research Challenge project TUDOR (Towards Ubiquitous 3D Open Resilient Network).}
    }  


\maketitle

\begin{abstract}
In this paper, we propose a \ac{PLS} framework for an \ac{IRS}-assisted \ac{ISASC} system, where a multi-antenna dual-functional semantic \ac{BS} serves multiple \acp{SCU} and monitors a potentially \ac{MST} in the presence of an \ac{EVE}. Both \ac{MST} and \ac{EVE} attempt to wiretap information from the signals transmitted to the \acp{SCU}.
The deployment of the \ac{IRS} not only enhances \ac{PLS} by directing a strong beam towards the \acp{SCU}, but also improves the localization information for the target without disclosing information about the \acp{SCU}.
To further strengthen \ac{PLS}, we employ joint \ac{AN} and \ac{DSS}, in addition to wiretap coding.
To evaluate sensing accuracy, we derive the \ac{CRB} for estimating the \ac{DoA}, and to assess the \ac{PLS} level of the \ac{ISASC} system, we determine a closed-form expression for the \ac{SSR}.
To achieve an optimal trade-off between these two competing objectives, we formulate a \ac{MOOP} for the joint design of the \ac{BS}'s \ac{BF} vectors and the \ac{IRS}'s phase shift vector.
To tackle this \ac{MOOP} problem, the $\epsilon$-constraint method is employed, followed by an alternating optimization (AO)-based algorithm that leverages the classical successive convex approximation (SCA) and semidefinite relaxation (SDR) techniques.
Simulation results demonstrate that the proposed scheme outperforms the baseline schemes, achieving a superior trade-off between \ac{SSR} and \ac{CRB}.
Specifically, our proposed approach improves the sensing accuracy by 5 dB compared to the commonly adopted maximal ratio transmission (MRT) approach.
\end{abstract}
\acresetall

\acresetall

\section{Introduction} \label{Introduction}
\IEEEPARstart{E}merging applications such as connected cars and smart factories expose the limitations of 5G infrastructure, demanding precise sensing, reliable and secure communication, and enhanced data processing. Integrating and co-designing sensing, communication, and computational tasks as fundamental services within a unified network architecture will be crucial for meeting these demands in beyond 5G and 6G \cite{Alwis2021}.

\Ac{ISAC}, which combines sensing and communication functions, has emerged as a key technology to improve energy and hardware efficiency \cite{T1, T2, T3, T4, T5, T6, T7, T8, T9, T10, T11, T12, T13, T14, Tnew}. Concurrently, the increasing data demands and spectrum scarcity have spurred interest in \ac{SC}. This AI-driven joint source-channel coding approach efficiently processes semantic information to alleviate network congestion \cite{T15, T16, T17, T18, T22}.

This paper proposes an \ac{ISASC} system that combines \ac{ISAC} and \ac{SC} strengths to optimize both sensing accuracy and data security, leveraging the mutual benefits from a co-design perspective.
However, integrating these technologies introduces significant security challenges.
Initially, due to the highly compressed nature of \ac{SC}, even minor data leaks can expose critical content, particularly when illegitimate users possess full background knowledge of the \acp{SCU} for semantic feature extraction. In addition, since transmitted signals contain both sensing and information data, they are susceptible to intercept.

As a result, \ac{ISASC} systems face two major security threats. The first is the traditional threat from \acs{EVE}, who aim to compromise communication confidentiality \cite{Tnew, T15, T16, T17, T18}. The second is a new threat introduced by ISAC systems: \Acs{MST}, which could exploit dual-use signals to access sensitive information \cite{T1, T2, T3, T4, T5, T6, T9, T10, T11, T12, T13, T14, Tnew}.

To address these threats, \ac{PLS} techniques, including techniques based on using \ac{AN} \cite{T3, T4, T5, T6, T8}, \ac{DSS} \cite{T2, T7, T11, T12, T13, T14, T22}, and \ac{IRS} \cite{T7, T8, T9, T10, T11, T12, T13, T14, Tnew, T18} have been developed. The goal of \ac{AN} is to ensure the confidentiality of data from \ac{EVE}, while \ac{DSS} aims to protect against exploitation by \ac{MST}, balancing the trade-off between secure transmission and effective sensing. 

Moreover, by adjusting the phase shifts of incident signals, \ac{IRS} can direct strong beams toward the \acp{SCU}, thus preventing interception attempts by \ac{EVE}.
In addition to enhancing \ac{PLS}, the use of \ac{IRS} is very appealing for \ac{ISASC} as it improves both communication and sensing performance. For communication, \ac{IRS} enhances the received signal strength at \acp{SCU}, improves channel ranks by providing additional signal paths, and reduces interference among \acp{SCU} \cite{T7, T8, T9}. For sensing, the \ac{IRS} enhances the accuracy of the location information of the targets in obstructed areas, while simultaneously reducing the amount of communication information revealed to \acp{MST} \cite{T10, T11}.
\\
\textbf{Related Works:}
In Table~\ref{Table_Overview}, we provide an overview of the literature related to secure \ac{ISAC}, \ac{SC}, and \ac{ISASC} systems.
On the one hand, we review the use of key performance indicators (KPIs) to evaluate sensing accuracy and communication confidentiality.
Specifically, for sensing accuracy, beam pattern gain (BPG) (equivalently, the \ac{SNR} of the echo signal) \cite{T2, T3, T4, T5, T8, T9, T11, T12, T13, T14, Tnew} and target detection probability (TDP) \cite{T7} are commonly used. Furthermore, the \ac{CRB} serves as a crucial KPI, providing a lower bound on the variance of unbiased parameter estimators \cite{T1, T6}.
For communication confidentiality, beyond using \ac{SNR} \cite{T9, T18} and \ac{SINR} \cite{T2, T7, T11, T12}, another line of work has adopted the \ac{SecR} as a fundamental metric to assess the effectiveness of \ac{PLS} frameworks \cite{T8, T1, T6, T3, T4, T5, T13, T14, Tnew}. In the semantic domain, \ac{SS} \cite{T17} is employed, and similar to \ac{SecR} in conventional communications, the \ac{SSR} \cite{T15, T16} is utilized as a metric for communication security level within this domain.

On the other hand, Table~\ref{Table_Overview} highlights key \ac{PLS} techniques employed in the studies to address various security challenges.
Specifically, some security issues for \ac{ISAC}, e.g., in the presence of \ac{EVE}, were explored in \cite{T7} and \cite{T8}, while the challenges posed by \ac{MST} were considered in \cite{T1, T2, T3, T4, T5, T6, T9, T10, T11, T12, T13, T14}. Notably, \cite{Tnew} investigates both \ac{EVE} and \ac{MST} threats jointly. Additionally, security problems associated with the presence of \ac{EVE} in \ac{SC} systems were addressed in \cite{T15, T16, T17, T18}.

However, these studies typically utilized either \ac{AN} or \ac{DSS} as \ac{PLS} techniques, along with various KPIs.
Moreover, despite the importance of designing secure systems, none of the previous work is focused on secure semantic \ac{ISAC}.
In this context, the joint utilizing \ac{AN} and \ac{DSS} strategies for \ac{IRS}-assisted integrated \ac{ISAC} and \ac{SC} systems is gaining renewed attention, driving the development of new transmission strategies to protect the exposure of private semantic information to both \ac{EVE} and \ac{MST} and achieve optimal security-aware \ac{ISASC} for future wireless networks.

\begin{table}[t] 
\caption{Overview of PLS frameworks in ISAC, SC, and ISASC Systems.}
\vspace{-.3cm}
\label{Table_Overview}
\centering
\small  
\setlength{\tabcolsep}{3pt}  
\renewcommand{\arraystretch}{1.1}  
\begin{tabular}{cccccccccc}
\hline
\multirow{2}{*}{\textbf{Ref.}} & \multirow{2}{*}{\textbf{SC}} & \multirow{2}{*}{\textbf{ISAC}} & \multicolumn{2}{c}{\textbf{KPIs}} & \multicolumn{2}{c}{\textbf{Threats}} & \multicolumn{3}{c}{\textbf{PLS}}\\ 
                 &       &       & \textbf{Com.}  &\textbf{Sen.}  &\textbf{EVE}&\textbf{MST}&\textbf{AN}& \textbf{DSS} & \textbf{IRS}  \\ \hline
\cite{T1}        & \ding{55} & \ding{51} & SecR             & CRB   & \ding{55} & \ding{51} & \ding{55} & \ding{55} & \ding{55} \\ 
\cite{T2}        & \ding{55} & \ding{51} & SINR           & BPG   & \ding{55} & \ding{51} & \ding{55} & \ding{51} & \ding{55} \\ 
\cite{T3,T4,T5}  & \ding{55} & \ding{51} & SecR             & BPG   & \ding{55} & \ding{51} & \ding{51} & \ding{55} & \ding{55} \\ 
\cite{T6}        & \ding{55} & \ding{51} & SecR             & CRB   & \ding{55} & \ding{51} & \ding{51} & \ding{55} & \ding{55} \\ 
\cite{T7}        & \ding{55} & \ding{51} & SINR           & TDP   & \ding{51} & \ding{55} & \ding{55} & \ding{51} & \ding{51} \\ 
\cite{T8}        & \ding{55} & \ding{51} & SecR             & BPG   & \ding{51} & \ding{55} & \ding{51} & \ding{55} & \ding{51} \\ 
\cite{T9}        & \ding{55} & \ding{51} & SNR            & BPG   & \ding{55} & \ding{51} & \ding{55} & \ding{55} & \ding{51} \\ 
\cite{T10}       & \ding{55} & \ding{51} & SumR           & CRB   & \ding{55} & \ding{51} & \ding{55} & \ding{55} & \ding{51} \\ 
\cite{T11,T12}   & \ding{55} & \ding{51} & SINR           & BPG   & \ding{55} & \ding{51} & \ding{55} & \ding{51} & \ding{51} \\ 
\cite{T13,T14}   & \ding{55} & \ding{51} & SecR             & BPG   & \ding{55} & \ding{51} & \ding{55} & \ding{51} & \ding{51} \\ 
\cite{Tnew}      & \ding{55} & \ding{51} & SecR             & BPG   & \ding{51} & \ding{51} & \ding{55} & \ding{55} & \ding{51} \\
\cite{T15,T16}   & \ding{51} & \ding{55} & SSR            & -     & \ding{51} & \ding{55} & \ding{55} & \ding{55} & \ding{55} \\ 
\cite{T17}       & \ding{51} & \ding{55} & SS             & -     & \ding{51} & \ding{55} & \ding{55} & \ding{55} & \ding{55} \\ 
\cite{T18}       & \ding{51} & \ding{55} & SNR            & -     & \ding{51} & \ding{55} & \ding{55} & \ding{55} & \ding{51} \\ 
\cite{T22}       & \ding{51} & \ding{51} & SSR            & BPG   & \ding{55} & \ding{51} & \ding{55} & \ding{51} & \ding{55} \\ \hline
\textbf{This paper} & \ding{51} & \ding{51} & SSR         & CRB   & \ding{51} & \ding{51} & \ding{51} & \ding{51} & \ding{51} \\ \hline
\end{tabular}
\end{table}
Specifically, \cite{T22} has explored secure resource allocation within \ac{ISASC} systems for the specific case of textual data. However, it neither provides a secure design against EVE, nor does it address the trade-off between \ac{SSR} and \ac{CRB}, nor does it unveil the role of the \ac{IRS} in enhancing \ac{PLS}.

Although our previous work \cite{WCNC2025} studied secure transmission designs for the ISASC system, it did not unveil the role of the IRS on security and sensing KPIs, and the proposed transceiver design is no longer applicable in the presence of an IRS.
\\
\textbf{Our Contributions:}
In this paper, to gain a better understanding of the interwoven sensing quality and communication confidentiality in the \ac{IRS}-assisted \ac{ISASC} system, we propose a \ac{MOOP} framework that sheds light on the trade-off between the \ac{CRB} and \ac{SSR}, two KPIs at the core of estimation theory and information theory.
Succinctly, the contributions of this paper can be summarized as follows:
\begin{itemize}
\item{We introduce a security-aware \ac{ISASC} system designed to mitigate security risks from both \ac{EVE} and \ac{MST} threats by incorporating joint \ac{AN} and \ac{DSS} as \ac{PLS} techniques.  
Furthermore, we explore using an \ac{IRS} to further enhance \ac{PLS}, simultaneously supporting \ac{SC} and sensing.}
\item{To evaluate sensing accuracy, we derive the \ac{CRB} for estimating the \ac{DoA} of a target, and to assess communication confidentiality, we determine a closed-form expression for the \ac{SSR}. Both metrics are calculated as functions of the communication, sensing, and \ac{AN} \ac{BF} vectors at the \ac{BS}, as well as the phase shift vector at the \ac{IRS}.} 
\item{We formulate a \ac{MOOP} for the joint design of the BS's \ac{BF} vectors and IRS's phase shift vector to achieve Pareto-optimal solutions that optimize the trade-off region between sensing and security.}
\item{By employing the $\epsilon$-constraint method, we reformulate the \ac{MOOP} into a \ac{SOOP}. To address the non-convex nature of this problem, we apply \ac{SDR}, \ac{SCA}, and \ac{GSS} techniques. Subsequently, we propose an algorithm to solve it based on alternating optimization, \ac{SDP}, and the \ac{GRM} methods.}
\item{Through numerical simulations, we validate that the proposed scheme achieves a larger trade-off region compared to baseline schemes.}
\end{itemize}
\textbf{Organization:}
In Section~\ref{System_Model}, we introduce the system and signal model, including the \ac{SC} transmitter, transmit signal model, communication model, sensing model and sensing KPI, \ac{EVE} model, and the \ac{SC} receiver along with the security KPI.
The \ac{MOOP} formulation and the proposed algorithm are discussed in Section~\ref{Problem_Formulation}.
Simulation results are presented in Section~\ref{Simulation_Results}.
Finally, conclusions are drawn in Section~\ref{Conclusion}.
\begin{figure*}[!t]
  \begin{center}
  \includegraphics[width=5.5in]{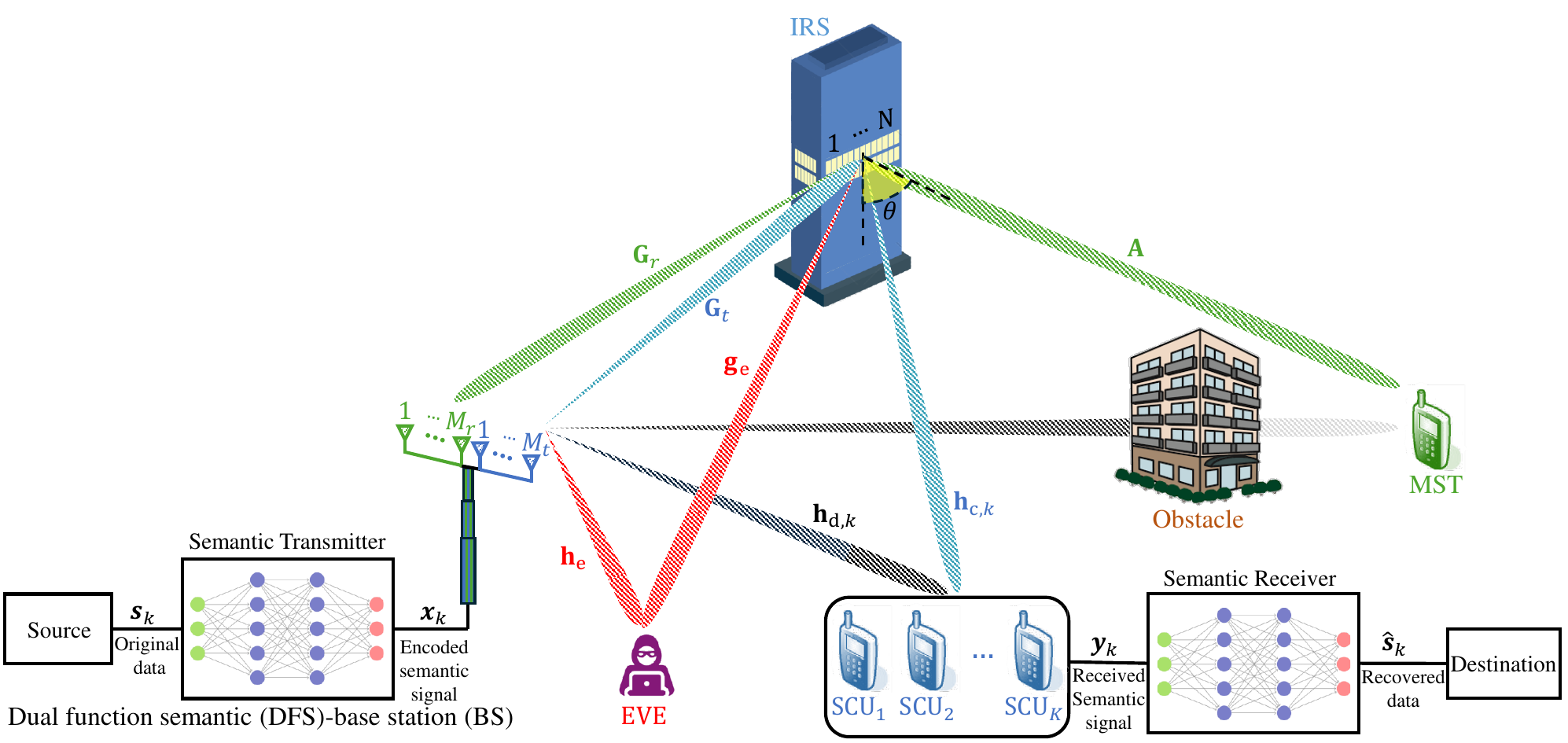}\\
  \caption{An IRS-assisted secure integrated sensing and semantic communication (ISASC) system.}\label{system_model}
  \end{center}
\end{figure*}
\\
\textbf{Notations:}
In this paper, for each complex number $a$, $|a|$ denotes its magnitude.
For each complex vector $\mathbf{a}$, $\mathbf{a}_i$ denotes the $i$-th element of $\mathbf{a}$, $\|\mathbf{a}\|$ denotes its Euclidean norm, $\mathbf{a}^\top$ denotes its transpose, and $\mathrm{diag}(\mathbf{a})$ denotes a diagonal matrix with each diagonal element being the corresponding element in $\mathbf{a}$.
For each complex matrix $\mathbf{A}$, $\mathbf{A}^{\dagger}$ denotes its conjugate transpose (or Hermitian), $\mathbf{A}^\ast$ denotes its element-wise complex conjugate, $\mathbf{A}^\top$ denotes its transpose. The operators $\mathbb{E}\{\mathbf{A}\}$ and $\operatorname{vec}(\mathbf{A})$ represent the statistical expectation and the vectorization of matrix $\mathbf{A}$, respectively. 
For a square complex matrix $\mathbf{A}$, $\operatorname{tr}(\mathbf{A})$ denotes its trace, and $\mathbf{A}^{-1}$ represents its inverse, respectively. The notation $\mathbf{A} \succcurlyeq \mathbf{0}$ means that $\mathbf{A}$ is a positive semi-definite matrix. 
The spaces of $n \times m$ real and complex matrices are represented by $\mathbb{R}^{n \times m}$ and $\mathbb{C}^{n \times m}$, respectively The distribution of a circularly symmetric complex Gaussian (CSCG) random variable with mean $a$ and variance $\sigma^2$ is denoted by $\mathcal{CN}(a, \sigma^2)$, and the symbol $\sim$ is used to denote ``distributed as." The real and imaginary parts of a complex number are denoted by $\operatorname{Re}\{\cdot\}$ and $\operatorname{Im}\{\cdot\}$, respectively.
$\mathcal{O}(.)$ denotes Landau's symbol, which describes the order of complexity. The matrix $\mathbf{I}_m$ is an identity matrix of dimension $m$. 


\section{System and Signal Model} \label{System_Model}
As illustrated in Fig.~\ref{system_model}, we consider a scenario involving a multi-antenna, dual-functional semantic BS equipped with $M_t$ transmit antennas and $M_r$ receive antennas. The BS serves $K$ single-antenna SCUs while simultaneously monitoring an MST, in the presence of a single-antenna EVE, with both the MST and EVE attempting to wiretap information from the SCUs' signals.
We assume that the direct links from the BS to the MST are obstructed. To mitigate this, an IRS composed of $N$ uniform linear array (ULA) reflecting elements is employed. The IRS assists in target sensing by creating a virtual line-of-sight (LoS) link and enhances PLS by directing strong beams from the BS to multiple legitimate SCUs.
The BS is further equipped with an SC transceiver, enabling it to effectively extract the underlying semantics of the transmitted data \cite{XIE21}.

\subsection {Semantic Communication Transmitter}
In this paper, we assume that the BS is equipped with an SC transceiver, allowing for effective extraction of the semantics underlying the source data. Specifically, the source generates data $\mathbf{s}_k = [s_k(1), s_k(2), \ldots, s_k(L_\text{s})]$, intended for the $k$-th SCU, where $s_k(i)$ denotes the $i$-th data segment, $i \in  \{1, \ldots, L_\text{s}\}$, and $L_\text{s}$ is the average number of data segments for $k$-th SCU, $k\in\mathcal{K}\triangleq\{1, \ldots, K\}$. The data is then processed by the semantic transmitter, which serves as a joint source-channel encoder \cite{XIE21}, and is mapped to a semantic symbol vector $\mathbf{x}_{\text{c},k} = [x_{\text{c},k}(1), x_{\text{c},k}(2), \ldots, x_{\text{c},k}(L)]$, where $\mathbf{x}_{\text{c},k}(l) ~\in~\mathbb{R}^{1 \times L}, l \in \mathcal{L} \triangleq \{1, \ldots, L\},$ represents the normalized power encoded semantic symbol and $L \triangleq \kappa L_\text{s}$ is the length of the semantic symbol vector. Here, $\kappa$ denotes the average number of encoded semantic symbols per data segment, serving as the scaling factor in the semantic transmitter that maps the length of the source data, $\mathbf{s}_k$, to the length of the encoded semantic symbols vector, $\mathbf{x}_{\text{c},k}$. 
Each semantic symbol can then be transmitted over the communication medium.

\subsection {Transmit Signal Model}
The BS transmits DSS as well as AN along with semantic symbol signals for communication, sensing, and anti-eavesdropping purposes. The combined transmit signal from the BS during the $l$-th symbol period can be expressed as
\begin{equation}\label{eq1}
\mathbf{x}(l) = \sum_{k \in \cal K}\mathbf{w}_{\text{c},k} x_{\text{c},k}(l) + \mathbf{w}_\text{s} x_\text{s}(l) + \mathbf{w}_\text{n} x_\text{n}(l),
\end{equation}
where $x_{\text{c},k}(l)$ represents the $l$-th encoded semantic symbol for SCU $k$, and $x_\text{s}(l)$ denotes the deterministic $l$-th DSS symbol, pre-coded with beamformer vector $\mathbf{w}_\text{s} \in \mathbb{C}^{M_t \times 1}$ to avoid leaking communication information to an MST and to enhance the localization information.
Additionally, $\mathbf{w}_{\text{c},k} \in \mathbb{C}^{M_t \times 1}$ denotes the communication transmit BF vector, and $x_\text{n}(l)$ stands for $l$-th AN signal, which follows the Gaussian distribution $x_\text{n}(l) \sim \mathcal{CN}(0, 1),\, l \in {\cal L}$, generated by BS and beamformed with $\mathbf{w}_\text{n} \in \mathbb{C}^{M_t \times 1}$ to avoid leaking information to the EVE.
We further assume that communication, sensing, and AN signals are statistically independent and uncorrelated.

We use the law of large numbers to approximate the sample covariance matrix of the transmitted signal $\mathbf{x}(l)$ with its statistical covariance matrix when the sample size $L$ is sufficiently large, i.e.,
\begin{align}\label{eq_2}
\mathbf{R}_x(\mathbf{w}) &\triangleq \frac{1}{L} \sum_{l \in \cal{L}}\mathbf{x}(l) \mathbf{x}^\dagger (l) \approx \mathbb{E}\{ \mathbf{x}(l) \mathbf{x}^\dagger (l) \} \\ \notag
             &= \sum_{k \in \cal K}\mathbf{W}_{\text{c},k} 
              + \mathbf{W}_\text{s} + \mathbf{W}_\text{n},
\end{align}
where $\mathbf{w}\in \mathbb{C}^{(K+2)M_t \times 1}$ denotes the BS BF vectors, defined as $\mathbf{w} \triangleq [\mathbf{w}_{\text{c},k}^\top, \mathbf{w}_\text{s}^\top, \mathbf{w}_\text{n}^\top]^\top,\ \forall k\in\mathcal{K}$.
Moreover, the matrices \(\mathbf{W}_{\text{c},k} = \mathbf{w}_{\text{c},k} \mathbf{w}_{\text{c},k}^\dagger\), \(\mathbf{W}_\text{s} = \mathbf{w}_\text{s} \mathbf{w}_\text{s}^\dagger\), and \(\mathbf{W}_\text{n} = \mathbf{w}_\text{n} \mathbf{w}_\text{n}^\dagger\) correspond to the covariance matrices for communication, sensing, and AN signals, respectively. 

\subsection {Communication Model}
We consider the scenario in which the IRS utilizes reflective beamforming to enhance ISASC operation, with the reflection phase shift vector and corresponding reflection matrix are denoted as $\mathbf{v} = [e^{j\phi_1}, \ldots, e^{j\phi_N}]^\top$ and $\mathbf{\Phi}(\mathbf{v}) = \operatorname{diag}(\mathbf{v})$, respectively. $v_n = e^{j\phi_n}$ is the phase shift of the $n$-th element at the IRS, $\phi_n\in[0,2\pi],\ n\in\mathcal{N}\triangleq \{1, \ldots, N\}$.
To simplify the notation, we will henceforth refer to $\mathbf{\Phi}(\mathbf{v})$ simply as $\mathbf{\Phi}$.

Let us define the composite channel between the BS and the $k$-th SCU as $\hat{\mathbf{h}}_{\text{Bc},k} \triangleq \mathbf{h}_{\text{c},k}^\dagger \mathbf{\Phi} \mathbf{G}_t + \mathbf{h}_{\text{d},k}^\dagger$, where $\mathbf{G}_t \in \mathbb{C}^{N \times M_t}$ is the channel matrix from the BS to the IRS. And $\mathbf{h}_{\text{d},k} \in \mathbb{C}^{M_t \times 1}$ and $\mathbf{h}_{\text{c},k} \in \mathbb{C}^{N \times 1}$ represent the channel vectors from the BS and the IRS to the $k$-th SCU, respectively. The signal received by the $k$-th SCU is expressed by
\begin{align}\label{eq3}
y^{\text{com}}_k(l) &= 
 \hat{\mathbf{h}}_{\text{Bc},k} \mathbf{w}_{\text{c},k} x_{\text{c},k}(l) 
&& \text{Desired semantic signal} \nonumber \\
& + \hat{\mathbf{h}}_{\text{Bc},k} \sum_{i\in \mathcal{K}, i\neq k} \mathbf{w}_{\text{c},i} x_{\text{c},i}(l) 
&& \text{Inter-user interference} \nonumber \\
& + \hat{\mathbf{h}}_{\text{Bc},k} \mathbf{w}_\text{s} x_\text{s}(l) 
&& \text{DSS interference} \nonumber \\
& + \hat{\mathbf{h}}_{\text{Bc},k} \mathbf{w}_\text{n} x_\text{n}(l) 
&& \text{AN interference} \nonumber \\
& + n_{\text{c},k}(l), 
&& \text{Receiver's noise}
\end{align}
where $n_{\text{c},k}(l) \sim \mathcal{CN}(0, \sigma_{\text{c},k}^2)$ denotes the noise at the $l$-th symbol period at the $k$-th SCU receiver. Consequently, the SINR at the $k$-th SCU can be expressed as,
\begin{align}\label{eq_4}
 &\gamma^\text{com}_k (\mathbf{w}, \mathbf{v})\\ \notag
 &=\frac{|\hat{\mathbf{h}}_{\text{Bc},k} \mathbf{w}_{\text{c},k}|^2}
                            {\sum_{i \in {\cal K},i \neq k}|\hat{\mathbf{h}}_{\text{Bc},k} \mathbf{w}_{\text{c},i}|^2 + |\hat{\mathbf{h}}_{\text{Bc},k} \mathbf{w}_\text{s}|^2 + |\hat{\mathbf{h}}_{\text{Bc},k} \mathbf{w}_\text{n}|^2 + \sigma_{\text{c},k}^2} \\ \notag
                      &=\frac{\hat{\mathbf{h}}_{\text{Bc},k} \mathbf{W}_{\text{c},k} \hat{\mathbf{h}}_{\text{Bc},k}^\dagger}
                            {\hat{\mathbf{h}}_{\text{Bc},k} (\mathbf{R}_x(\mathbf{w}) - \mathbf{W}_{\text{c},k}) \hat{\mathbf{h}}_{\text{Bc},k}^\dagger + \sigma_{\text{c},k}^2}.
\end{align}

\subsection {Sensing Model and Sensing KPI}
Given that the direct link between the BS and the MST is obstructed, the IRS is utilized to establish a virtual LoS link for effective sensing. The received echo signal at the BS receiver is described by
\begin{equation}\label{eq_5}
\mathbf{y}^\text{echo}(l) = (\mathbf{G}_r \mathbf{\Phi}^\top \mathbf{A} \mathbf{\Phi} \mathbf{G}_t) \mathbf{x}(l) + \mathbf{n}_\text{s}(l),
\end{equation}
where $\mathbf{G}_r \in \mathbb{C}^{M_r \times N}$ denotes the channel matrix from the IRS to the BS\footnote{It should be noted that $\mathbf{G}_r=\mathbf{G}_t^\top$ for the monostatic multiple-input multiple-output (MIMO) system, with $M_t = M_r$ \cite{GUO15}.} and $\mathbf{n}_\text{s}(l) \sim \mathcal{CN}(0, {\sigma_\text{s}}^2 \mathbf{I}_{M_r})$ represents the noise at the BS, which may include environmental clutter. In (\ref{eq_5}), $\mathbf{A}$ denotes the IRS-MST-IRS cascaded echo channel, defined as
\begin{equation}\label{eq_6}
\mathbf{A} = \alpha \mathbf{a}(\theta) \mathbf{a}^\top(\theta),
\end{equation}
where $\alpha \in \mathbb{C}^1$ is the complex-valued channel coefficient influenced by the target radar cross-section (RCS) and the round-trip path loss of the echo link at the IRS.\footnote{It is assumed that the target is sufficiently small so that the incident signal is reflected from a single point on its surface.} Additionally, the steering vector $\mathbf{a}(\theta)$ at the IRS, associated with the angle $\theta$, is given by
\begin{equation}\label{eq_7}
\mathbf{a}(\theta) = \begin{bmatrix}
1, e^{j \frac{2\pi d_\text{IRS}}{\lambda} \sin \theta}, \dots, e^{j \frac{2\pi (N-1) d_\text{IRS}}{\lambda} \sin \theta}
\end{bmatrix}^\top,
\end{equation}
where $\theta$ denotes the target’s DoA at the IRS, $d_\text{IRS}$ specifies the spacing between adjacent IRS elements, and $\lambda$ is the carrier wavelength.

For simplicity, we assume that the radar dwell time corresponds to the length of the semantic symbol vector, i.e., $L$, and we stack the transmitted signals, the received signals, and the noise over $L$ as $\mathbf{X} = [\mathbf{x}(1), \ldots, \mathbf{x}(L)]$, $\mathbf{Y}^\text{echo} = [\mathbf{y}^\text{echo}(1), \ldots, \mathbf{y}^\text{echo}(L)]$, and $\mathbf{N}_\text{s} = [\mathbf{n}_\text{s}(1), \ldots, \mathbf{n}_\text{s}(L)]$, respectively. Accordingly, we have
\begin{equation}\label{eq_8}
\mathbf{Y}^\text{echo} = \alpha \mathbf{H}_\text{BB}(\mathbf{v}) \mathbf{X} + \mathbf{N}_\text{s},
\end{equation}
where, for the sake of conciseness, we define $\mathbf{H}_\text{BB}(\mathbf{v}) \triangleq \mathbf{c}(\mathbf{v}) \mathbf{b}^\top(\mathbf{v})$, with $\mathbf{b}(\mathbf{v}) = \mathbf{G}_t^\top \mathbf{\Phi}^\top \mathbf{a}(\theta)$ and $\mathbf{c}(\mathbf{v}) = \mathbf{G}_r \mathbf{\Phi}^\top \mathbf{a}(\theta)$. 
For notation convenience, in the following, we will omit $\mathbf{v}$ and simply denote $\mathbf{H}_\text{BB}(\mathbf{v})$, $\mathbf{b}(\mathbf{v})$, and $\mathbf{c}(\mathbf{v})$ as $\mathbf{H}_\text{BB}$, $\mathbf{b}$, $\mathbf{c}$, respectively.
By vectorizing (\ref{eq_8}), we obtain
\begin{equation}\label{eq_9}
\tilde{\mathbf{y}} = \operatorname{vec}(\mathbf{Y}^\text{echo}) = \tilde{\mathbf{x}} + \tilde{\mathbf{n}},
\end{equation}
where $\tilde{\mathbf{x}} = \alpha \operatorname{vec}(\mathbf{H}_\text{BB} \mathbf{X})$ and $\tilde{\mathbf{n}} = \operatorname{vec}(\mathbf{N}_\text{s})$ follows a CSCG distribution, $\mathcal{CN}(0, {\sigma_\text{s}}^2 \mathbf{I}_{{M_r}L})$.

Define the vector of unknown parameters as $\mathbf{\xi} = [\theta, \bm{\alpha}^\top]^\top \in \mathbb{R}^{3 \times 1}$, with $\bm{\alpha} = [\alpha_\text{Re}, \alpha_\text{Im}]^\top$ denoting the real and imaginary parts of the channel coefficient, specifically $\alpha_\text{Re} = \operatorname{Re}\{\bm{\alpha}\}$ and $\alpha_\text{Im} = \operatorname{Im}\{\bm{\alpha}\}$. 
The Fisher information matrix (FIM) for the estimation of these unknown parameters, $\bm{\xi}$, is represented as
\begin{equation}\label{eq_10}
\mathbf{F}_{\bm{\xi}} = \begin{bmatrix}
f_{\theta\theta} & \mathbf{f}_{\theta\bm{\alpha}} \\
\mathbf{f}_{\theta\bm{\alpha}}^\top & \mathbf{f}_{\bm{\alpha}\bm{\alpha}}
\end{bmatrix} \in \mathbb{R}^{3 \times 3}.
\end{equation}
where \cite[eq. (10)]{SONG23},
\begin{align}
f_{\theta\theta} &= \frac{2L |\alpha|^2}{{\sigma_\text{s}}^2} \operatorname{tr}(\dot{\mathbf{H}}_\text{BB} \mathbf{R}_x(\mathbf{w}) \dot{\mathbf{H}}_\text{BB}^\dagger), \label{eq_12a}\\ 
\mathbf{f}_{\theta\bm{\alpha}} &= \frac{2L}{{\sigma_\text{s}}^2} \operatorname{Re} \{\alpha^* \operatorname{tr}(\mathbf{H}_\text{BB} \mathbf{R}_x(\mathbf{w}) \dot{\mathbf{H}}_\text{BB}^\dagger) [1, j]\}, \label{eq_12b}\\ 
\mathbf{f}_{\bm{\alpha}\bm{\alpha}} &= \frac{2L}{{\sigma_\text{s}}^2} \operatorname{tr}(\mathbf{H}_\text{BB} \mathbf{R}_x(\mathbf{w}) \mathbf{H}_\text{BB}^\dagger) \mathbf{I}_2, \label{eq_12c}
\end{align}
with $\dot{\mathbf{H}}_\text{BB} = \dot{\mathbf{c}} \mathbf{b}^\top + \mathbf{c} \dot{\mathbf{b}}^\top$ represents the partial derivative of $\mathbf{H}_\text{BB}$ with respect to (w.r.t.) $\theta$, and $\dot{\mathbf{b}} = j \frac{2\pi d_\text{IRS}}{\lambda} \cos \theta \, \mathbf{G}_t^\top \mathbf{\Phi}^\top \mathbf{D}\,\mathbf{a}(\theta)$ and $\dot{\mathbf{c}} = j \frac{2\pi d_\text{IRS}}{\lambda} \cos \theta \, \mathbf{G}_r \mathbf{\Phi}^\top \mathbf{D}\,\mathbf{a}(\theta)$ denote the partial derivative of $\mathbf{b}$ and $\mathbf{c}$ w.r.t. $\theta$, respectively, with $\mathbf{D} = \operatorname{diag}(0, 1, \ldots, N-1)$. 

In this paper, we focus on characterizing the CRB for estimating the target’s DoA, i.e., $\operatorname{CRB}_\theta$, which corresponds to the first diagonal element of the inverse FIM, given by:
\begin{equation}\label{eq_13}
\operatorname{CRB}_\theta = \mathbf{F}_{\bm{\xi}}^{-1}(1,1) = (f_{\theta\theta} - \mathbf{f}_{\theta\bm{\alpha}} \mathbf{f}_{\bm{\alpha}\bm{\alpha}}^{-1} \mathbf{f}_{\theta\bm{\alpha}}^\top)^{-1}.
\end{equation}
Thus, by inserting (\ref{eq_12a}) - (\ref{eq_12c}) into (\ref{eq_13}), the CRB for the DoA, $\theta$, can be explicitly expressed as a function of the BS communication, sensing, and AN BF vectors, as well as the IRS phase shift vector, as
\begin{align}\label{eq_14}
\operatorname{CRB}_\theta(\mathbf{w}, \mathbf{v})
= \frac{{{\sigma_\text{s}}^2}}{2L |\alpha|^2 \,J(\mathbf{w}, \mathbf{v})},
\end{align}
where $J(\mathbf{w}, \mathbf{v})\triangleq\operatorname{tr}(\dot{\mathbf{H}}_\text{BB} \mathbf{R}_x(\mathbf{w}) \dot{\mathbf{H}}_\text{BB}^\dagger) - \frac{|\operatorname{tr}(\mathbf{H}_\text{BB} \mathbf{R}_x(\mathbf{w}) \dot{\mathbf{H}}_\text{BB}^\dagger)|^2}{\operatorname{tr}(\mathbf{H}_\text{BB} \mathbf{R}_x(\mathbf{w}) \mathbf{H}_\text{BB}^\dagger)}$. Which (\ref{eq_14}) provides a quantifiable measure, showing the lower bound on the variance of any unbiased estimator of $\theta$, incorporating the effects of the BS's BF strategies, i.e., $\mathbf{w}$, and the IRS configuration, i.e., $\mathbf{v}$.

\subsection {Eavesdropper Model}
We define the composite channel between the BS and the EVE as \(\hat{\mathbf{h}}_\text{Be} \triangleq \mathbf{g}_\text{e}^\dag \mathbf{\Phi} \mathbf{G}_t + \mathbf{h}_\text{e}^\dag\), where \(\mathbf{h}_\text{e} \in \mathbb{C}^{M_t \times 1}\) and \(\mathbf{g}_\text{e} \in \mathbb{C}^{N \times 1}\) are the channel vectors from the BS and the IRS to the EVE, respectively. The signal received by the EVE is given by
\begin{equation}\label{eq_15}
y^\text{eve}(l) = \hat{\mathbf{h}}_\text{Be} \mathbf{x}(l) + n_\text{e}(l),
\end{equation}
where \(n_\text{e}(l) \sim \mathcal{CN}(0, \sigma_\text{e}^2)\) represents the noise experienced at the EVE at symbol $l$. The SINR at the EVE when attempting to intercept the information of the $k$-th SCU is derived as,
\begin{equation}\label{eq_16}
\gamma^\text{eve}_k(\mathbf{w}, \mathbf{v}) = \frac{\hat{\mathbf{h}}_{\text{Be},k} \mathbf{W}_{\text{c},k} \hat{\mathbf{h}}_{\text{Be}}^\dagger}
                           {\hat{\mathbf{h}}_{\text{Be}} (\mathbf{R}_x(\mathbf{w}) - \mathbf{W}_{\text{c},k}) \hat{\mathbf{h}}_{\text{Be}}^\dagger + \sigma_{\text{e}}^2}.
\end{equation}

\subsection {KPI for Security and Reliability}
To assess the performance of SC, we utilize the semantic similarity between the original data, $\mathbf{s}_k$, and its reconstructed version from the semantic symbols, $\hat{\mathbf{s}}_k$, as the KPI \cite{XIE21}. The semantic similarity ranges from 0 to 1, where a higher value indicates a higher degree of similarity between $\mathbf{s}_k$ and $\hat{\mathbf{s}}_k$.
Recently, \cite{MU23} introduced a tractable function for semantic similarity, employing a generalized logistic function based on the average number of encoded semantic symbols per data segment, \(\kappa\), (i.e., the employed semantic encoding/decoding scheme) and the SINR of the received signal, \(\gamma\),
\begin{equation}\label{eq_18}
\mathcal{E}(\kappa, \gamma) = A_{\kappa,1} + \frac{A_{\kappa,2} - A_{\kappa,1}}{1 + \exp{(-C_{\kappa,1}(10 \log{\gamma}) - C_{\kappa,2})}},
\end{equation}
where \(A_{\kappa,1} > 0\) and  \(A_{\kappa,2}> 0\) represent the lower (left) and upper (right) asymptotes, respectively, while \(C_{\kappa,1}> 0\) indicates the logistic growth rate, and \(C_{\kappa,2}\) determines the midpoint of the logistic function, as adopted in \cite[eq.~(3)]{MU23}.\footnote{The logistic function's broad functionality allows this metric to extend to different types of sources, including text, speech, images, and video by setting appropriate approximation parameters \cite{Liu2024, mashhad}.}

Assuming $I$ represents the average semantic information carried by the source message, $\mathbf{s}_k$ (measured in semantic units, or \textit{suts}), the semantic information per symbol is ${I}/{\kappa L_\text{s}}$ (\textit{suts/sym}). Given that the symbol rate matches the channel bandwidth in passband transmission, the achievable semantic rate (SR) in \textit{suts/sec} for the $k$-th SCU (or EVE for wiretapping the information of the $k$-th SCU\footnote{Note that our scenario is a worst-case scenario for this setup as we consider full side information at EVE about the SCUs' background knowledge.}) over a channel with bandwidth $B$ can be expressed as \cite[eq. (4)]{YAN23}
\begin{equation}\label{eq_19}
\operatorname{SR}^\chi_k (\mathbf{w}, \mathbf{v})= \frac{BI}{\kappa L_\text{s}} \mathcal{E}(\kappa, \gamma^\chi_k(\mathbf{w}, \mathbf{v})), \chi\in\{\text{com},\text{eve}\}.
\end{equation}
Consequently, the SSR of the $k$-th SCU can be given by 
\begin{align}\label{eq_21}
\operatorname{SSR}_k(\mathbf{w}, \mathbf{v}) &= [\operatorname{SR}^\text{com}_k(\mathbf{w}, \mathbf{v}) - \operatorname{SR}^\text{eve}_k(\mathbf{w}, \mathbf{v})]^+,
\end{align}
where \([ \cdot ]^+ \triangleq \max(0, \cdot)\) \cite[eq.~(17)]{T15}, \cite[eq.~(10)]{T22}.

\section {Multi-Objective Optimization Design and Proposed scheme} \label{Problem_Formulation}
From \eqref{eq_14} and \eqref{eq_21}, it is evident that both CRB and SSR depend significantly on the BF vectors at the BS, i.e., $\mathbf{w}$, and the phase shift vector at the IRS, i.e., $\mathbf{v}$.\footnote{It is assumed that the channel state information (CSI) for both SCUs and EVE is perfectly known at the BS. Specifically, the CSI for SCUs can be obtained using efficient channel estimation methods \cite{channel1}, while the CSI for EVE can be acquired through CSI feedback methods \cite{channel2} or by detecting EVE's local oscillator leakage power \cite{channel3}.} Therefore, in this section, we present an MOOP to optimize the BS's BF vectors and the IRS's phase shift vector, balancing CRB and SSR. Following this, we complete our proposed security-aware IRS-assisted ISASC framework by specifying appropriate algorithms that focus on the sensing-security trade-off.

\subsection{Problem Formulation}
To comprehensively investigate the trade-off region between CRB and SSR, we adopt an MOOP to simultaneously optimize these two objectives. Unlike SOOPs, which focus on a single metric, MOOP can generate a diverse set of Pareto-optimal solutions that address both objectives effectively \cite{NGA05}.
Considering the system model of Section \ref{System_Model}, our goal is to develop an MOOP that simultaneously minimizes the CRB and maximizes the minimum SSR by finding the optimal communication, sensing, and AN BF vectors at the BS, i.e., $\mathbf{w}\triangleq [\mathbf{w}_{\text{c},k}^\top, \mathbf{w}_\text{s}^\top, \mathbf{w}_\text{n}^\top]^\top,\ \forall k\in\mathcal{K}$, as well as phase shift vector at the IRS, i.e., $\mathbf{v}$. This MOOP can be mathematically formulated as,
\begin{subequations}
\begin{align}
(\cal {P}):\ & \underset{\mathbf{w}\in \mathbb{C}^{(K+2)M_t\times1}, \mathbf{v}\in \mathbb{C}^{N\times1}}{\Min} \ {\rm CRB}_\theta(\mathbf{w}, \mathbf{v}), \label{eqP_a} \\
& \underset{\mathbf{w}, \mathbf{v}}{\Max} \min_{k \in \mathcal{K}} \operatorname{SSR}_k(\mathbf{w}, \mathbf{v}), \label{eqP_b} \\
\text{s.t.} \quad &\operatorname{tr}(\mathbf{R}_x) \le P_{\max}, \label{eqP_c}\\
&|v_n| \le 1, \ \forall n \in \mathcal{N}, \label{eqP_d}
\end{align}
\end{subequations}
where $P_{\max}$ in \eqref{eqP_c} represents the maximum transmit power at the BS. Furthermore, each IRS phase shift is required to be unit-modulus, as specified in constraint \eqref{eqP_d}.

\subsection{Problem Reformulation}
To solve the MOOP, we employ the $\epsilon$-constraint method by transferring max-min SSR into a constraint, making the CRB minimization the primary objective function \cite{CHI13}. Notably, the $\epsilon$-constraint method can generate the entire Pareto frontier of the two objective metrics by varying the value of $\epsilon$ and solving the corresponding optimization problem. Consequently, for a given $\epsilon \geq 0$, the reformulated optimization problem to obtain a Pareto-optimal solution can be expressed as
\begin{subequations}
\begin{align}
(\mathcal{P}.1):\ &\underset{\mathbf{w}, \mathbf{v}}{\Min} \ {\rm CRB}_\theta(\mathbf{w}, \mathbf{v}), \label{eqP1_a} \\
\text{s.t.} \quad & \min_{k \in \mathcal{K}} \operatorname{SSR}_k(\mathbf{w}, \mathbf{v}) \geq \epsilon, \label{eqP1_b} \\
&\eqref{eqP_c}, \eqref{eqP_d}, \nonumber 
\end{align}
\end{subequations}
where constraint \eqref{eqP1_b} guarantees the minimum SSR is larger than \(\epsilon\).
Note that the operator $[ \cdot ]^+$ can be removed without altering the value of \eqref{eq_21} because the SSR can always be guaranteed to be non-negative via constraint \eqref{eqP1_b}. Thus, the problem ($\mathcal{P}.1$) becomes
\begin{subequations}
\begin{align}
(\mathcal{P}.2):\ &\underset{\mathbf{w}, \mathbf{v}}{\Min} \ {\rm CRB}_\theta(\mathbf{w}, \mathbf{v}), \label{eqP22_a} \\
\text{s.t.} \quad &\operatorname{SR}_k^\text{com}(\mathbf{w}, \mathbf{v}) - \operatorname{SR}^\text{eve}_k(\mathbf{w}, \mathbf{v}) \geq \epsilon, \ \forall k \in \mathcal{K}, \label{eqP22_b} \\
&\eqref{eqP_c}, \eqref{eqP_d}. \nonumber 
\end{align}
\end{subequations}

Next, by introducing the slack variable \(r_\text{th}\), we can reformulate the problem (\(\mathcal{P}.2\)) to an equivalent problem as
\begin{subequations}
\begin{align}
(\mathcal{P}.3):\ &\underset{\mathbf{w}, \mathbf{v}, \,r_\text{th}\in \mathbb{R}^{1 \times 1}}{\Min} \ {\rm CRB}_\theta (\mathbf{w}, \mathbf{v}), \label{eqP3_a} \\
\text{s.t.} \quad & \operatorname{SR}_k^\text{com}(\mathbf{w}, \mathbf{v}) \geq r_\text{th}+ \epsilon , \ \forall k \in \mathcal{K}, \label{eqP3_b} \\
& \operatorname{SR}^\text{eve}_k(\mathbf{w}, \mathbf{v}) \leq r_\text{th}, \ \forall k \in \mathcal{K}, \label{eqP3_c} \\
& \eqref{eqP_c}, \eqref{eqP_d}, \nonumber
\end{align}
\end{subequations}
where $r_\text{th}$ denotes the maximum allowable information leakage to the EVE’s link.
By incorporating expressions \eqref{eq_19} into constraints \eqref{eqP3_b} and \eqref{eqP3_c}, we redefine the semantic rate constraints in problem $(\mathcal{P}.3)$ as
\begin{align}
\gamma^\text{com}_k (\mathbf{w}, \mathbf{v}) \geq \Gamma_\text{th}^\text{com} & \triangleq 10^{\frac{-1}{10 C_{\kappa,1}} (C_{\kappa,2} + \ln{\left[\frac{A_{\kappa,2} - \tilde{r}_\text{th}^\text{com}}{\tilde{r}_\text{th}^\text{com} - A_{\kappa,1}}\right]})}, \label{eq_28} \\ 
\gamma^\text{eve}_k (\mathbf{w}, \mathbf{v}) \leq \Gamma_\text{th}^\text{eve} & \triangleq 10^{\frac{-1}{10 C_{\kappa,1}} (C_{\kappa,2} + \ln{\left[\frac{A_{\kappa,2} - \tilde{r}_\text{th}^\text{eve}}{\tilde{r}_\text{th}^\text{eve} - A_{\kappa,1}}\right]})}, \label{eq_29}
\end{align}
where $\tilde{r}_\text{th}^\text{com} \triangleq ( r_\text{th} + \epsilon)L/BI$ and $\tilde{r}_\text{th}^\text{eve} \triangleq r_\text{th}L/BI$. Thus, the problem $(\mathcal{P}.3)$ can be reformulated as,
\begin{subequations}
\begin{align}
(\mathcal{P}.4):\ &\underset{\mathbf{w}, \mathbf{v}, \,r_\text{th}}{\Min} \ {\rm CRB}_\theta (\mathbf{w}, \mathbf{v}), \label{eqP4_a} \\
\text{s.t.} \quad & \gamma_k^\text{com} (\mathbf{w}, \mathbf{v}) \geq \Gamma_\text{th}^\text{com}, \ \forall k \in \mathcal{K}, \label{eqP4_b} \\
& \gamma^\text{eve}_k (\mathbf{w}, \mathbf{v}) \leq \Gamma_\text{th}^\text{eve}, \ \forall k \in \mathcal{K}, \label{eqP4_c} \\
& \eqref{eqP_c}, \eqref{eqP_d}. \nonumber
\end{align}
\end{subequations}
Problem ($\mathcal{P}.4$) can be reformulated into two optimization problems: a single-variable optimization problem on $r_\text{th}$, and an optimization problem concerning the BS BF vectors and the IRS phase shift vector, as follows:
\begin{subequations}
\begin{align}
(\mathcal{P}.5.1):\ & \underset{r_\text{th}}{\Min} \ {\cal F}(r_\text{th}) \label{eqP5_a} \\
\text{s.t.} \quad & \frac{BI}{L}A_{\kappa,1} \le r_\text{th}
\le\frac{BI}{L}A_{\kappa,2}-\epsilon,\label{eqP5_b} 
\end{align}
\end{subequations}
with
\begin{subequations}
\begin{align}
(\mathcal{P}.5.2):\ {\cal F}(r_\text{th}) \triangleq \ & \underset{\mathbf{w}, \mathbf{v}}{\Min} \ {\rm CRB}_\theta(\mathbf{w}, \mathbf{v}) \label{eqP6_a} \\
\text{s.t.} \quad & \eqref{eqP4_b}, \eqref{eqP4_c}, \eqref{eqP_c}, \eqref{eqP_d}. \nonumber
\end{align}
\end{subequations}
where the inequalities in \eqref{eqP5_b} follow from \eqref{eq_28} and \eqref{eq_29}, ensuring the logarithm's argument remains positive. Since $r_\text{th}$ lies in the interval $[(BI/L)A_{\kappa,1}, (BI/L)A_{\kappa,2}-\epsilon]$, the single-variable optimization problem $(\mathcal{P}.5.1)$ can be addressed by derivative-free search algorithms for solving one-dimensional optimization problems, such as GSS \cite{BER97}, which is summarized in Algorithm~\ref{Algorithm1}.

\begin{algorithm}[t!]
\caption{Proposed algorithm for problem~($\mathcal{P}$).}
\begin{algorithmic}
\STATE
\STATE 1:  \textbf{Initialization:} 
\STATE \hspace{.4cm}Define the SSR search regime $\epsilon \in [\epsilon_\text{min},\epsilon_\text{max}]$.
\STATE \hspace{.4cm}Define \(\delta_{\text{GSS}}\) as the convergence accuracy threshold.
\STATE \hspace{.4cm}Initialize \(\mathbf{w}\) and \(\mathbf{v}\).
\STATE \hspace{.4cm}Initialize parameters for \( r_\text{th} \):
\STATE \hspace{.6cm} i. Set \( a = (BI/L)A_{\kappa,1}\) and \( b = (BI/L)A_{\kappa,2}-\epsilon \).
\STATE \hspace{.6cm}ii. Define the reduction ratio \( \tau = \frac{\sqrt{5} - 1}{2} \).
\STATE 2:  \textbf{For} each candidate $\epsilon$, set constraint \eqref{eqP1_b}, \textbf{do:}
\STATE 3:  \textbf{Repeat} until \( |b - a| \leq \delta_\text{GSS} \):
\STATE \hspace{.6cm}   i.   Calculate \( c = (b - \tau) (b - a) \) and \( d = (a + \tau) (b - a) \).
\STATE \hspace{.6cm}  ii.  Evaluate \( \mathcal{F}(c) \) and \( \mathcal{F}(d) \) by solving the (\(\mathcal{P}.5.2\)).
\STATE \hspace{.6cm} iii. If \( \mathcal{F}(c) < \mathcal{F}(d) \), set \( b = d \); otherwise, set \( a = c \).
\STATE 4:  \textbf{Set:} The optimal value \( r^\text{opt}_{\text{th}} = \frac{a + b}{2} \).
\STATE 5:  \textbf{Evaluate} \( \mathcal{F}(r^\text{opt}_{\text{th}}) \) by solving the (\(\mathcal{P}.5.2\)).
\STATE 6:  \textbf{Output:} The optimal \(\mathbf{w}^\text{opt}\) and \(\mathbf{v}^\text{opt}\) for the current \(\epsilon\).
\end{algorithmic}\label{Algorithm1}
\end{algorithm}

After these transformations, it becomes evident that the original problem $(\mathcal{P})$ can be effectively decomposed, allowing us to focus more specifically on the refined challenge presented in problem $(\mathcal{P}.5.2)$.
However, $(\mathcal{P}.5.2)$ is still NP-hard due to the coupling of variables. To address this, in the subsequent subsections, we will decompose $(\mathcal{P}.5.2)$ into two sub-problems and solve it using the AO approach. Each sub-problem focuses on optimizing specific variables while keeping the others fixed. Specifically, we alternately optimize the BS BF vectors $\mathbf{w}$ and the IRS phase shift vector $\mathbf{v}$ through the following two related sub-problems:\\ 
\textit{Sub-problem~1}: Optimization of BS BF vectors, i.e., $\mathbf{w}$, with a given IRS phase shift vector, i.e., $\mathbf{v}$.\\
\textit{Sub-problem~2}: Optimization of IRS phase shift vector, i.e., $\mathbf{v}$, with given BS BF vectors, i.e., $\mathbf{w}$.\\
The advantage of this decomposition is that each sub-problem is a convex SDP, which can be optimally solved using convex optimization solvers such as CVX in MATLAB \cite{BOYD04}. 

We then present the overall algorithm in subsection~\ref{Overall} and investigate its convergence and complexity in subsection~\ref{Complexity}.

\subsection {Sub-Problem 1: Optimization of BS BF Vectors} \label{Subproblem1}
According to \eqref{eq_14}, minimizing ${\rm CRB}_\theta (\mathbf{w}, \mathbf{v})$ can be equivalently rewritten to maximize $J(\mathbf{w}, \mathbf{v})$.
Accordingly, by fixing the IRS phase shift vector $\mathbf{v}$, we formulate sub-problem~1 $(\mathcal{SP}1)$ as,
\begin{subequations}
\begin{align}
(\mathcal{SP}1): \ & \underset{\mathbf{w}}{\Max} \ J(\mathbf{w}, \mathbf{v}) \label{SP1_a}\\
\text{s.t.} \quad & \eqref{eqP4_b}, \eqref{eqP4_c}, \eqref{eqP_c}.\nonumber
\end{align}
\end{subequations}

By introducing an auxiliary variable $t$, applying the Schur's complement \cite{FUZ05}, $(\mathcal{SP}1)$ is re-expressed as,\footnote{To simplify the notation, we will henceforth refer to $\mathbf{R}_x(\mathbf{w})$ as $\mathbf{R}_x$.}
\begin{subequations}
\begin{align}
&(\mathcal{SP}1.1):\ \underset{\mathbf{w},\,t\in\mathbb{R}^{1\times1}}{\Max} \ t \label{SP1_2_a}\\
&\text{s.t.} \left[
\begin{matrix}
\operatorname{tr}(\dot{\mathbf{H}}_\text{BB} \mathbf{R}_x \dot{\mathbf{H}}_\text{BB}^\dagger) - t & \operatorname{tr}(\mathbf{H}_\text{BB} \mathbf{R}_x \dot{\mathbf{H}}_\text{BB}^\dagger) \\
\operatorname{tr}(\dot{\mathbf{H}}_\text{BB} \mathbf{R}_x \mathbf{H}_\text{BB}^\dagger) & \operatorname{tr}(\mathbf{H}_\text{BB} \mathbf{R}_x \mathbf{H}_\text{BB}^\dagger)
\end{matrix}
\right] \succcurlyeq \mathbf{0}, \label{SP1_2_b}\\
& \quad \ \ \eqref{eqP4_b}, \eqref{eqP4_c}, \eqref{eqP_c}.\nonumber
\end{align}
\end{subequations}

The constraints in $(\mathcal{SP}1.1)$ involve quadratic variables, i.e., \(\mathbf{W}_{\text{c},k}\), \(\mathbf{W}_\text{s}\), and \(\mathbf{W}_\text{n}\), making it challenging to solve. To facilitate the solution, $(\mathcal{SP}1.1)$ is reformulated based on BF matrices, ${{\mathbf{W}}} \triangleq [\mathbf{W}_{\text{c},k},\mathbf{W}_\text{s},\mathbf{W}_\text{n}], \forall k \in \mathcal{K}$:
\begin{subequations}
\begin{align}
&(\mathcal{SP}1.2): \ \underset{{\mathbf{W}\in \mathbb{C}^{M_t\times(K+2)M_t}},\,t}{\Max} \ t \label{SP1_3_a} \\ 
& \text{s.t.} \nonumber \\ 
& {\mathbf{h}}_{\text{Bc},k} \big((1 + \frac{1}{\Gamma_{k}^\text{com}})\mathbf{W}_{\text{c},k} - \mathbf{R}_x\big) {\mathbf{h}}_{\text{Bc},k}^\dagger \geq \sigma_{\text{c},k}^2, \ \forall k \in \mathcal{K}, \label{SP1_1_c} \\
& {\mathbf{h}}_\text{Be} \big((1 + \frac{1}{\Gamma_{k}^\text{eve}}) \mathbf{W}_{\text{c},k} - \mathbf{R}_x\big) {\mathbf{h}}_\text{Be}^\dagger \leq \sigma_\text{e}^2, \ \forall k \in \mathcal{K}, \label{SP1_1_d} \\
& \mathbf{W}_{\text{c},k} \succcurlyeq \mathbf{0}, \ \mathbf{W}_\text{s} \succcurlyeq \mathbf{0}, \ \mathbf{W}_\text{n} \succcurlyeq \mathbf{0}, \ \forall k \in \mathcal{K}, \label{SP1_1_g} \\
& \operatorname{rank}(\mathbf{W}_{\text{c},k}) = 1, \operatorname{rank}(\mathbf{W}_\text{s}) = 1, \operatorname{rank}(\mathbf{W}_\text{n}) = 1, \forall k \in \mathcal{K},\label{SP1_1_h} \\
& \eqref{SP1_2_b}, \eqref{eqP_c}.\nonumber
\end{align}
\end{subequations}

The sole non-convexity in solving $(\mathcal{SP}1.2)$ arises from the rank-one constraints in \eqref{SP1_1_h}. By relaxing these constraints, $(\mathcal{SP}1.1)$ can be transformed into an SDR problem, which can be efficiently solved using convex optimization solvers \cite{BOYD04}.
After obtaining the optimal BS BF matrices, ${{\mathbf{W}}}^\text{opt}$, we employ the GRM to construct the rank-one solution, i.e.,  $\mathbf{w}^{\text{opt}}$ \cite{LUO10}.

\subsection {Sub-Problem 2: IRS Phase Shift Vector Optimization} \label{Subproblem2}
In sub-problem~2 $(\mathcal{SP}2)$, with fixed BS BF vectors, optimization focuses solely on the IRS phase shift vector. Thus, problem $(\mathcal{P}.5.2)$ is reduced to
\begin{subequations}
\begin{align}
(\mathcal{SP}2):\ & \underset{\mathbf{v}}{\Min} \ {\rm CRB}_\theta (\mathbf{w}, \mathbf{v}) \label{eqSP_2_a} \\
\text{s.t.} \quad & \eqref{eqP4_b},\eqref{eqP4_c},\eqref{eqP_d}. \nonumber
\end{align}
\end{subequations}

To address this optimization problem, we first re-express the SINR constraints \eqref{eqP4_b} and \eqref{eqP4_c} w.r.t \(\mathbf{v}\). Introducing \(\widetilde{\mathbf{G}}_{\text{c},k} \triangleq \operatorname{diag}(\mathbf{h}_{\text{c},k}^\dagger) \mathbf{G}_t\) allows us to express \(\mathbf{h}_{\text{c},k}^\dagger \mathbf{\Phi} \mathbf{G}_t = \mathbf{v}^\top \widetilde{\mathbf{G}}_{\text{c},k}\). Consequently, we can rewrite the SINR for the $k$-th SCU, from \eqref{eq_4}, w.r.t. \(\mathbf{v}\) as 
\begin{equation}\label{eq:31}
\gamma^\text{com}_k(\mathbf{w}, \mathbf{v}) = \frac{ \widetilde{\mathbf{v}}^\dagger \hat{\mathbf{G}}_{\text{c},k} \mathbf{W}_{\text{c},k}^\ast \hat{\mathbf{G}}_{\text{c},k}^\dagger \widetilde{\mathbf{v}}}
{\widetilde{\mathbf{v}}^\dagger \hat{\mathbf{G}}_{\text{c},k} (\mathbf{R}_x^\ast - \mathbf{W}_{\text{c},k}^\ast) \hat{\mathbf{G}}_{\text{c},k}^\dagger \widetilde{\mathbf{v}} + \sigma_{\text{c},k}^2},
\end{equation}
where \(\widetilde{\mathbf{v}} \triangleq [\mathbf{v}^\top, 1]^\top\) and \(\hat{\mathbf{G}}_{\text{c},k} \triangleq [\widetilde{\mathbf{G}}_{\text{c},k}^\dagger, \mathbf{h}_{\text{d},k}]^\top\).
Based on \eqref{eq:31}, we have transformed the SINR constraint in \eqref{eqP4_b} w.r.t. \(\mathbf{v}\) as
\begin{equation}\label{eq:32}
\widetilde{\mathbf{v}}^\dagger \mathbf{C}^\text{com}_k \widetilde{\mathbf{v}} \geq \Gamma_{k}^\text{com} \sigma_{\text{c},k}^2,
\end{equation}
where $\mathbf{C}^\text{com}_k$ is defined as $\mathbf{C}^\text{com}_k \triangleq \hat{\mathbf{G}}_{\text{c},k} ((1+\Gamma_{k}^\text{com})\mathbf{W}_{\text{c},k}^\ast - \Gamma_{k}^\text{com} \mathbf{R}_x^\ast) \hat{\mathbf{G}}_{\text{c},k}^\dagger$.
 Similarly, the second SINR constraint from \eqref{eqP4_c} is transformed w.r.t. $\mathbf{v}$ as
\begin{equation}\label{eq:33}
\widetilde{\mathbf{v}}^\dagger \mathbf{C}^\text{eve}_k \widetilde{\mathbf{v}} \leq \Gamma_{k}^\text{eve} \sigma_\text{e}^2,
\end{equation}
where \(\mathbf{C}^\text{eve}_k \triangleq \hat{\mathbf{G}}_\text{e} ((1+\Gamma_{k}^\text{eve})\mathbf{W}_{\text{c},k}^\ast - \Gamma_{k}^\text{eve} \mathbf{R}_x^\ast) \hat{\mathbf{G}}_\text{e}^\dagger\), with \(\hat{\mathbf{G}}_\text{e} = [\widetilde{\mathbf{G}}_\text{e}^\dagger, \mathbf{h}_\text{e}]^\top\) and \(\widetilde{\mathbf{G}}_\text{e} \triangleq \operatorname{diag}(\mathbf{g}_\text{e}^\dagger) \mathbf{G}_t\).

Next, we re-express the objective function in $(\mathcal{SP}2)$ w.r.t. \(\mathbf{v}\). By introducing \(\widetilde{\mathbf{A}} \triangleq \operatorname{diag}(\mathbf{a}(\theta))\), we can express \(\mathbf{b} = \mathbf{G}_t^\top \widetilde{\mathbf{A}} \mathbf{v}\), \(\mathbf{c} = \mathbf{G}_r \widetilde{\mathbf{A}} \mathbf{v}\), \(\dot{\mathbf{b}} = j 2\pi \frac{d_\text{IRS}}{\lambda} \cos{\theta}\, \mathbf{G}_t^\top \widetilde{\mathbf{A}} \mathbf{D} \mathbf{v}\), and \(\dot{\mathbf{c}} = j 2\pi \frac{d_\text{IRS}}{\lambda} \cos{\theta}\, \mathbf{G}_r \widetilde{\mathbf{A}} \mathbf{D} \mathbf{v}\), with $\mathbf{D} \triangleq \operatorname{diag}(0, 1, \cdots, N - 1)$. Consequently,
\begin{equation}\label{eq:28}
\mathbf{H}_\text{BB} = \mathbf{c} \mathbf{b}^\top = \mathbf{G}_r \widetilde{\mathbf{A}} \mathbf{v} \mathbf{v}^\top \widetilde{\mathbf{A}}^\top \mathbf{G}_t,
\end{equation}
and
\begin{align}\label{eq:29}
&\dot{\mathbf{H}}_\text{BB} = \dot{\mathbf{c}} \mathbf{b}^\top + \mathbf{c} \dot{\mathbf{b}}^\top \nonumber\\
&\ \quad = j 2\pi \frac{d_\text{IRS}}{\lambda} \cos{\theta} \, \mathbf{G}_r \widetilde{\mathbf{A}} (\mathbf{D} \mathbf{v} \mathbf{v}^\top + \mathbf{v} \mathbf{v}^\top \mathbf{D}^\top) \widetilde{\mathbf{A}}^\top \mathbf{G}_t.
\end{align}

By substituting \eqref{eq:28} and \eqref{eq:29} into \eqref{eq_14}, the objective function in $(\mathcal{SP}2)$ can be re-expressed w.r.t. \(\mathbf{v}\) as
\begin{equation} \label{eq_Const_CRB_v}
J(\mathbf{w}, \mathbf{v}) = \frac{J_1(\mathbf{v})+J_2(\mathbf{v})}{{{(4 \pi^2 d_\text{IRS}^2 \cos^2{\theta})}/\lambda^2}},
\end{equation} 
where
\begin{equation} J_{i_1}(\mathbf{v}) \triangleq \mathbf{v}^\dagger \mathbf{R}_{i_2} \mathbf{v} ( \mathbf{v}^\dagger \mathbf{D} \mathbf{R}_{i_1} \mathbf{D} \mathbf{v} - \frac{| \mathbf{v}^\dagger \mathbf{D} \mathbf{R}_{i_1} \mathbf{v} |^2}{\mathbf{v}^\dagger \mathbf{R}_{i_1} \mathbf{v}} ), \label{eq_f_general} \end{equation}
with $(i_1,i_2)\in \{(1,2), (2,1)\}$, and we define \(\mathbf{R}_1 \triangleq \widetilde{\mathbf{A}}^\dagger \mathbf{G}_r^\dagger \mathbf{G}_r \widetilde{\mathbf{A}}\) and \(\mathbf{R}_2 \triangleq \widetilde{\mathbf{A}}^\dagger \mathbf{G}_t^\ast \mathbf{R}_x^\ast \mathbf{G}_t^\top \widetilde{\mathbf{A}}\).

Finally, by substituting \eqref{eq:32}, \eqref{eq:33}, and \eqref{eq_Const_CRB_v} into $(\mathcal{SP}2)$, the IRS phase shift vector optimization problem is transformed as
\begin{subequations}
\begin{align}
(\mathcal{SP}2.1):\ & \underset{\mathbf{v}}{\Max}\ J_1(\mathbf{v}) + J_2(\mathbf{v})\label{SP2_1_a} \\
\text{s.t.} \quad & \eqref{eq:32},\eqref{eq:33},\eqref{eqP_d}. \nonumber \label{SP2_1_b}
\end{align}
\end{subequations}
Problem $(\mathcal{SP}2.1)$ is non-convex due to the non-concave unit-modulus constraint in \eqref{eqP_d} and the objective function in \eqref{SP2_1_a}. To address these challenges, we apply SDR to the former and SCA to the latter.
Initially, we define \(\mathbf{V} = \widetilde{\mathbf{v}} \widetilde{\mathbf{v}}^\dagger\) as the SDR auxiliary variable, with \(\mathbf{V} \succcurlyeq \mathbf{0}\) and \(\operatorname{rank}(\mathbf{V}) = 1\). To satisfy constraint \eqref{eqP_d}, we consider \(\mathbf{V}_{n,n} = 1\) for \(n \in \mathcal{N}\). Then, using the following transformations w.r.t. \(\mathbf{V}\): $\mathbf{v}^\dagger \mathbf{R}_{\hat{i}} \mathbf{v} = \operatorname{tr}(\widetilde{\mathbf{R}}_{\hat{i}} \mathbf{V})$, $\mathbf{v}^\dagger \mathbf{D} \mathbf{R}_{\hat{i}} \mathbf{v} = \operatorname{tr}(\widetilde{\mathbf{D}} \widetilde{\mathbf{R}}_{\hat{i}} \mathbf{V})$, $\mathbf{v}^\dagger \mathbf{D} \mathbf{R}_{\hat{i}} \mathbf{D} \mathbf{v} = \operatorname{tr}(\widetilde{\mathbf{D}} \widetilde{\mathbf{R}}_{\hat{i}} \widetilde{\mathbf{D}} \mathbf{V})$, $\hat{i}\in\{1,2\}$, and $\widetilde{\mathbf{v}}^\dagger \mathbf{C}^\chi_k \widetilde{\mathbf{v}} = \operatorname{tr}(\mathbf{C}^\chi_k \mathbf{V}), \chi\in\{\text{com},\text{eve}\}$, where
\[
\widetilde{\mathbf{R}}_{\hat{i}} = \begin{bmatrix} \mathbf{R}_{\hat{i}} & \mathbf{0}_{N \times 1} \\ \mathbf{0}_{1 \times N} & 0 \end{bmatrix}, \quad \widetilde{\mathbf{D}} = \begin{bmatrix} \mathbf{D} & \mathbf{0}_{N \times 1} \\ \mathbf{0}_{1 \times N} & 0 \end{bmatrix},
\]
$J_{i_1}(\mathbf{V})$ in \eqref{eq_f_general} can be reformulated as 
\begin{equation} 
\tilde{J}_{i_1}(\mathbf{V})=\operatorname{tr}(\widetilde{\mathbf{R}}_{i_2} \mathbf{V})(  \operatorname{tr}(\widetilde{\mathbf{D}} \widetilde{\mathbf{R}}_{i_1} \widetilde{\mathbf{D}} \mathbf{V}) - \frac{| \operatorname{tr}(\widetilde{\mathbf{D}} \widetilde{\mathbf{R}}_{i_1} \mathbf{V})|^2}{\operatorname{tr}(\widetilde{\mathbf{R}}_{i_1} \mathbf{V})}).
\end{equation} 

Finally, $(\mathcal{SP}2.1)$ can be reformulated w.r.t. \(\mathbf{V}\) as
\begin{subequations}
\begin{align}
(\mathcal{SP}2.2):\ &\underset{\mathbf{V}\in\mathbb{C}^{N\times N}}{\Max}\ \tilde{J}_1(\mathbf{V})+\tilde{J}_2(\mathbf{V})\label{SP2_2_a}\\
\text{s.t.} \quad &\operatorname{tr}(\mathbf{C}^\text{com}_k \mathbf{V}) \geq \Gamma^\text{com}_k \sigma_{\text{c},k}^2, \ \forall k \in \mathcal{K}, \label{SP2_2_b} \\
&\operatorname{tr}(\mathbf{C}^\text{eve}_k \mathbf{V}) \leq \Gamma^\text{eve}_k \sigma_\text{e}^2, \ \forall k \in \mathcal{K}, \label{SP2_2_c}\\
& \mathbf{V} \succcurlyeq \mathbf{0},\ \mathbf{V}_{n,n} = 1, \ n \in \mathcal{N}, \label{SP2_2_e}\\
& \operatorname{rank}(\mathbf{V}) = 1. \label{SP2_2_f}
\end{align}
\end{subequations}
Although the rank-one constraint in \eqref{SP2_2_f} is non-convex, we can drop it to obtain the SDR version of (\(\mathcal{SP}2.2\)). 

To handle the non-concave objective function in \eqref{SP2_2_a}, we introduce a new slack variable $\mathbf{u}\triangleq [u_1, u_2]^\top$, and re-express problem ($\mathcal{SP}2.2$) with these auxiliary variables as
\begin{subequations}
\begin{align}
(\mathcal{SP}2.3):\ &\underset{\mathbf{V}, \mathbf{u}\in \mathbb{R}^{2 \times 1}}{\Max}\ \bar{J}_1(\mathbf{V},\mathbf{u})+\bar{J}_2(\mathbf{V},\mathbf{u})\label{SP2_3_a}\\
\text{s.t.} \quad &\frac{| \operatorname{tr}(\widetilde{\mathbf{D}} \widetilde{\mathbf{R}}_{\hat{i}} \mathbf{V}) |^2}{\operatorname{tr}(\widetilde{\mathbf{R}}_{\hat{i}} \mathbf{V})} \leq u_{\bar{i}}, \ {\hat{i}}\in \{1,2\} \label{SP2_3_b} \\
& \eqref{SP2_2_b}-\eqref{SP2_2_e}, \nonumber
\end{align}
\end{subequations}
where \(\bar{J}_1(\mathbf{V}, \mathbf{u})\) is a convex function w.r.t. \(\mathbf{V}\), defined as
\begin{align}\label{eq:47}
\bar{J}_1(\mathbf{V}, \mathbf{u}) &= \frac{1}{4} ( \operatorname{tr}(\widetilde{\mathbf{R}}_2 \mathbf{V}) + \operatorname{tr}(\widetilde{\mathbf{D}} \widetilde{\mathbf{R}}_1 \widetilde{\mathbf{D}} \mathbf{V}) )^2 \nonumber \\
& + \frac{1}{4} ( \operatorname{tr}(\widetilde{\mathbf{R}}_1 \mathbf{V}) + \operatorname{tr}(\widetilde{\mathbf{D}} \widetilde{\mathbf{R}}_2 \widetilde{\mathbf{D}} \mathbf{V}))^2 \nonumber \\
& + \frac{1}{4} ( \operatorname{tr}(\widetilde{\mathbf{R}}_2 \mathbf{V}) - u_1)^2 \nonumber \\
& +\frac{1}{4} ( \operatorname{tr}(\widetilde{\mathbf{R}}_1 \mathbf{V}) - u_2 )^2,
\end{align}
and \(\bar{J}_2(\mathbf{V}, \mathbf{u})\) is a concave function w.r.t. \(\mathbf{V}\), defined as
\begin{align}\label{eq:48}
\bar{J}_2(\mathbf{V}, \mathbf{u}) = &-\frac{1}{4} ( \operatorname{tr}(\widetilde{\mathbf{R}}_2 \mathbf{V}) - \operatorname{tr}(\widetilde{\mathbf{D}} \widetilde{\mathbf{R}}_1 \widetilde{\mathbf{D}} \mathbf{V}))^2 \nonumber \\
&- \frac{1}{4} ( \operatorname{tr}(\widetilde{\mathbf{R}}_1 \mathbf{V}) - \operatorname{tr}(\widetilde{\mathbf{D}} \widetilde{\mathbf{R}}_2 \widetilde{\mathbf{D}} \mathbf{V}))^2 \nonumber \\
&- \frac{1}{4} ( \operatorname{tr}(\widetilde{\mathbf{R}}_2 \mathbf{V}) + u_1)^2 \nonumber \\
&- \frac{1}{4} ( \operatorname{tr}(\widetilde{\mathbf{R}}_1 \mathbf{V}) + u_2)^2.
\end{align}
Using Schur’s complement \cite{FUZ05}, the constraint in \eqref{SP2_3_b} can be transformed into convex semidefinite constraints as
\begin{align}
\left[ \begin{matrix} u_{\hat{i}} & \operatorname{tr}(\widetilde{\mathbf{D}} \widetilde{\mathbf{R}}_{\hat{i}} \mathbf{V}) \\ \operatorname{tr}(\mathbf{V}^\dagger \widetilde{\mathbf{R}}_{\hat{i}}^\dagger \widetilde{\mathbf{D}}^\dagger) & \operatorname{tr}(\widetilde{\mathbf{R}}_{\hat{i}} \mathbf{V}) \end{matrix} \right] \succcurlyeq \mathbf{0}, \ \hat{i}\in \{1, 2\} \label{eq:47a}
\end{align}

However, (\(\mathcal{SP}2.3\)) remains non-convex due to the non-concave nature of \(\bar{J}_1(\mathbf{V}, \mathbf{u})\) in the objective function. 
To address this, we employ the SCA approach. Specifically, in the $r$-th iteration of the SCA algorithm, \(\bar{J}_1(\mathbf{V}, \mathbf{u})\) can be approximated with a linear function as
\begin{align}\label{eq:51}
\hat{J}_1^{(r)}(\mathbf{V}, \mathbf{u}) &\triangleq \bar{J}_1(\mathbf{V}^{(r)}, \mathbf{u}^{(r)}) \nonumber \\
&+ \frac{1}{2} \operatorname{tr} ( (\widetilde{\mathbf{R}}_2 + \widetilde{\mathbf{D}} \widetilde{\mathbf{R}}_1 \widetilde{\mathbf{D}}) \mathbf{V}^{(r)}) \nonumber \\ 
&\times \operatorname{tr} ( (\widetilde{\mathbf{R}}_2 + \widetilde{\mathbf{D}} \widetilde{\mathbf{R}}_1 \widetilde{\mathbf{D}}) (\mathbf{V} - \mathbf{V}^{(r)})) \nonumber \\
&+ \frac{1}{2} \operatorname{tr} ( (\widetilde{\mathbf{R}}_1 + \widetilde{\mathbf{D}} \widetilde{\mathbf{R}}_2 \widetilde{\mathbf{D}}) \mathbf{V}^{(r)})
\nonumber \\
&\times \operatorname{tr} ( (\widetilde{\mathbf{R}}_1 + \widetilde{\mathbf{D}} \widetilde{\mathbf{R}}_2 \widetilde{\mathbf{D}}) (\mathbf{V} - \mathbf{V}^{(r)})) \nonumber \\
&+ \frac{1}{2} ( u_1^{(r)} - \operatorname{tr} (\widetilde{\mathbf{R}}_2 \mathbf{V}^{(r)})) ( u_1 - u_1^{(r)}) \nonumber \\
&+ \frac{1}{2} ( u_2^{(r)} - \operatorname{tr} (\widetilde{\mathbf{R}}_1 \mathbf{V}^{(r)})) ( u_2 - u_2^{(r)}),
\end{align}
which is the first-order Taylor expansion of \(\bar{J}_1(\mathbf{V}, \mathbf{u})\) around the local solution \((\mathbf{V}^{(r)}, \mathbf{u}^{(r)})\).

By replacing \(\bar{J}_1(\mathbf{V}, \mathbf{u})\) with \(\hat{J}_1^{(r)}(\mathbf{V}, \mathbf{u})\), (\(\mathcal{SP}2.3\)) can be approximated as the following convex problem at the $r$-th iteration of the SCA:
\begin{subequations}
\begin{align}
(\mathcal{SP}2.4):\ &\underset{\mathbf{V}, \mathbf{u}}{\Max}\ \hat{J}_1^{(r)}(\mathbf{V}, \mathbf{u}) + \bar{J}_2(\mathbf{V}, \mathbf{u}) \label{SP2_4_a} \\
\text{s.t.} \quad & \eqref{eq:47a}, \eqref{SP2_2_b} - \eqref{SP2_2_f} \nonumber.
\end{align}
\end{subequations}
This convex problem can be optimally solved by convex solvers such as CVX.
Finally, based on the optimal solution \(\mathbf{V}^\text{opt}\) from (\(\mathcal{SP}2.4\)), we apply GRM to construct an approximate rank-one solution \cite{LUO10}.

\subsection {Proposed Algorithm for Problem (\(\mathcal{P}.5.2\))} \label{Overall}
This section summarizes the proposed AO-SDP algorithm for addressing the non-convex NP-hard optimization problem (\(\mathcal{P}.5.2\)), which involves the joint optimization of the BF vectors $\mathbf{w}$ and phase shift vector $\mathbf{v}$. This problem, due to its complexity and non-convex nature, has been decomposed into more manageable sub-problems that are solvable using convex optimization techniques, specifically (\(\mathcal{SP}1.3\)) for BS BF vectors optimization and (\(\mathcal{SP}2.4\)) for IRS phase shift vector optimization. We provide a detailed description of the proposed algorithm for solving (\(\mathcal{P}.5.2\)) in Algorithm~\ref{Algorithm2}.

\begin{algorithm}[t!]
\caption{Proposed AO-SDP-based algorithm for problem~($\mathcal{P}.5.2$).}
\begin{algorithmic}
\STATE
\STATE 1:  \textbf{Initialization:}
\STATE \hspace{.4cm}The AO iteration index $t=0$.
\STATE \hspace{.4cm}Randomly generate an initial phase shift vector $\mathbf{v}^{(0)}$.
\STATE 2:  \textbf{Repeat:}
\STATE 3:  \hspace{0.5cm}Increment $t=t+1$.
\STATE \textit{\textbf{Sub-problem 1: BF vectors optimization}}
\STATE 4:  \hspace{0.5cm}With the given ${\mathbf{v}^{(t-1)}}$, perform the BF matrices,
\STATE \hspace{.8cm}${\mathbf{W}}^{(t)}$, by solving the relaxed SDP in (\(\mathcal{SP}1.3\)).
\STATE 5:  \hspace{0.5cm}Construct rank-one solution using the GRM, $\mathbf{w}^{(t)}$.
\STATE \textit{\textbf{Sub-problem 2: Phase shift vector optimization}}
\STATE 6:  \hspace{0.5cm}Initialize the SCA iteration index $r = 0$, set $\mathbf{V}^{(0)}=$
\STATE \hspace{.9cm}$\mathbf{v}^{(t-1)}{(\mathbf{v}^{(t-1)})}^\dagger$.
\STATE 7:  \hspace{0.5cm}\textbf{Repeat:}
\STATE 8:  \hspace{.8cm}i. Increment $r = r+1$.
\STATE 9:  \hspace{.7cm}ii. Based on the current $\mathbf{V}^{(r-1)}$, compute interm-
\STATE \hspace{1.6cm}ediate variables $u^{(r-1)}_1 = \frac{| \operatorname{tr}(\widetilde{\mathbf{D}} \widetilde{\mathbf{R}}_1 \mathbf{V}^{(r-1)})|^2}{\operatorname{tr}(\widetilde{\mathbf{R}}_1 \mathbf{V}^{(r-1)})}$,
\STATE \hspace{1.6cm}$u^{(r-1)}_2 = \frac{| \operatorname{tr}(\widetilde{\mathbf{D}} \widetilde{\mathbf{R}}_2 \mathbf{V}^{(r-1)})|^2}{\operatorname{tr}(\widetilde{\mathbf{R}}_2 \mathbf{V}^{(r-1)})}$.
\STATE 10:  \hspace{.7cm}iii. Construct function $\hat{J}_1^{(r-1)}(\mathbf{V}^{(r-1)}, \mathbf{u}^{(r-1)})$.
\STATE 11:  \hspace{.7cm}iv. With the current BF matrices ${\mathbf{w}}^{(t)}$,
\STATE \hspace{1.6cm} calculate the IRS phase shift matrix $\mathbf{V}^{(r)}$ by
\STATE \hspace{1.6cm} solving (\(\mathcal{SP}2.4\)).
\STATE 12:  \hspace{0.5cm}\textbf{Until:} The maximum number of SCA iterations, 
\STATE \hspace{1cm}$N_\text{SCA}$, is reached.
\STATE 13:  \hspace{0.5cm}Construct rank-one solution for the phase shift vector
\STATE \hspace{1cm}$\mathbf{v}^{(t)}$ using GRM.
\STATE 14:  \textbf{Until:}  Meets the convergence threshold \(\delta_{\text{AO}}\).
\STATE 15: \textbf{Output:} $\mathbf{w}^{\text{opt}}=[{{\mathbf{w}_{\text{c},k}^\top}^{\text{opt}}}, {{\mathbf{w}_\text{s}^\top}^{\text{opt}}}, {{\mathbf{w}_\text{n}^\top}^{\text{opt}}}]^\top$, $\mathbf{v}^{\text{opt}}$.
\end{algorithmic}\label{Algorithm2}
\end{algorithm}

\subsection {Convergence and Computational Complexity Analysis} \label{Complexity}
In this section, we analyze the convergence and complexity of the overall algorithm.
\subsubsection {Convergence Analysis}
The convergence of the proposed algorithm is guaranteed for the following reasons:
\begin{itemize}{}{}
\item{Sub-problem (\(\mathcal{SP}1\)) and (\(\mathcal{SP}2\)) are solved optimally during each iteration, ensuring that the objective function value does not decrease.}
\item{With a sufficient number of GRM, the SDR approach achieves at least a \(\pi/4\)-approximation of the optimal objective value \cite{LUO10}.\footnote{Note that the rank-one solution from the GRM might decrease the objective function. To ensure algorithm convergence, a large number of GRM iterations are required to maintain a non-decreasing objective function at each step \cite{LI23}.} This ensures that the objective function value generally increases or remains constant. If the objective function decreases, the iteration is terminated.}
\item{Both objective functions, i.e., CRB and SSR, are bounded within the feasible set of (\(\mathcal{P}\)). This ensures that the values cannot become infinite.}
\end{itemize}
Although global optimality cannot be claimed due to the non-convex nature of problem (\(\mathcal{P}\)), the algorithm is guaranteed to achieve at least a locally optimal solution \cite{HONG16}.

\subsubsection {Computational Complexity Analysis}
According to \cite{complexity1}, the computational complexity of Algorithm~\ref{Algorithm2} is primarily driven by solving SDP problems in two sub-problems iteratively: sub-problem 1 - optimization of BS beamforming (BF) vectors (\(\mathcal{SP}1.3\)) in step~4, and sub-problem 2 - IRS phase shift vector optimization (\(\mathcal{SP}2.4\)) in step~11.
Specifically, the complexity of solving an SDP problem involving a set of $\mathsf{n}$ SDP constraints, each containing a $\mathsf{m}\times \mathsf{m}$ positive semidefinite matrix, is given by $\mathcal{O}(\mathsf{n}\mathsf{m}^3 + \mathsf{n}^2\mathsf{m}^2 + \mathsf{n}^3)$ \cite[Theorem 3.12]{complexity2}. 
For the BS BF vectors optimization (\(\mathcal{SP}1.3\)), where $\mathsf{n} = 2K+1$ and $\mathsf{m} = (K+2)M_\text{t}$, the computational complexity is $\mathcal{O}_{\mathcal{SP}1}((2K+1)((K+2)M_\text{t})^3 + (2K+1)^2((K+2)M_\text{t})^2 + (2K+1)^3)\approx\mathcal{O}_{\mathcal{SP}1}(K^4 M_\text{t}^3)$.
Similarly, for the IRS phase shift vector optimization SCA problem (\(\mathcal{SP}2.4\)), with $\mathsf{n} = 2K+N$ and $\mathsf{m} = N$, the complexity is given by $\mathcal{O}_{\mathcal{SP}2}(N_\text{SCA}[(2K+N)N^3 + (2K+N)^2N^2 + (2K+N)^3]) \approx \mathcal{O}_{\mathcal{SP}2}(N^4)$, where $N_\text{SCA}$ denotes the maximum number of iterations in the SCA algorithm.
Thus, the total complexity of Algorithm~\ref{Algorithm2} is $\mathcal{O}_\text{AO}(\log(\frac{1}{\delta_\text{AO}})[\mathcal{O}_{\mathcal{SP}1}+\mathcal{O}_{\mathcal{SP}2}])$, where $\delta_\text{AO} > 0$ is the predefined accuracy factor for Algorithm~\ref{Algorithm2}.
Assuming that the range of $\epsilon$ is divided into $N_\epsilon$ discrete values, the overall computational complexity of the proposed algorithm in Algorithm~\ref{Algorithm1} is $\mathcal{O}( N_\epsilon \log(\frac{1}{\delta_\text{GSS}}) \mathcal{O}_\text{AO})\approx \mathcal{O}(K^4 M_\text{t}^3 +N^4) $, where $\delta_\text{GSS}> 0$ represents the accuracy factor for Algorithm~\ref{Algorithm1}.

\renewcommand\arraystretch{1.3}
\begin{table}[!t]
\caption{System Parameters for BS, IRS, SCUs, MST, and EVE.}
\centering
\begin{tabular}{p{22pt}p{152pt}p{30pt}}
\hline
\bf{Symbol} &\bf{Description} &\bf{Value}\\
\hline
\( M_t \) & Number of transmit antennas in the BS& 6-8 \\
\( M_r \) & Number of receive antennas in the BS& 6-8 \\
\( N \) & Number of reflecting elements in the IRS & 8-32 \\
\( K \) & Number of SCUs & 2 \\
\( L_\text{s} \) & Length of the data segment & 256 \\
\( \sigma_\text{s}^2 \) & Noise power at the BS & -90~dBm \\
\( \sigma_{\text{c},k}^2 \) & Noise power at the SCUs & -90~dBm \\
\( \sigma_\text{e}^2 \) & Noise power at the EVE & -90~dBm \\
\( d_{\text{IRS}} \) & Spacing between adjacent reflecting elements & \( \lambda/2 \) \\
\( B \) & Bandwidth & 5 MHz \\
\( I \) & Average amount of semantic information  & 10 \\
\( P_{\max} \) & Maximum transmit power at the BS & 30~dBm \\
 \hline
\label{Table_1}
\end{tabular}
\end{table}

\section{Simulation Results} \label{Simulation_Results}
This section provides numerical results to evaluate the performance of the proposed algorithm detailed in Sections~\ref{Problem_Formulation}.
For each transmission path, we model the conventional distance-dependent path loss as $L(d) = K_0 ({d}/{d_0})^{-\zeta_0}$, where \(d\) is the distance of the transmission link and \(K_0 = -30 \, \text{dB}\) is the path loss at the reference distance \(d_0 = 1 \, \text{m}\) \cite{Hamid_Nomadic}. The path loss exponent \(\zeta_0\) is set as 2.5 for IRS-assisted links (i.e., \(\mathbf{G}_t\), \(\mathbf{G}_r\), \(\mathbf{h}_{\text{c},k}\), and \(\mathbf{g}_\text{e}\)) and 3.5 for the direct links (i.e., \(\mathbf{h}_{\text{d},k}\) and \(\mathbf{h}_\text{e}\)) \cite{LI24}. For IRS-assisted links, Rician fading is assumed
\begin{equation}
\mathbf{G}_\tau = \sqrt{\frac{\beta_\text{BI}}{1+\beta_\text{BI}}} \mathbf{G}_\tau^{\text{LoS}} + \sqrt{\frac{1}{1+\beta_\text{BI}}} \mathbf{G}_\tau^{\text{NLoS}}, \tau\in \{t, r\} \label{eq:53_1}
\end{equation}
\begin{equation}
\mathbf{h}_{\text{c},k} = \sqrt{\frac{\beta_\text{IC}}{1+\beta_\text{IC}}} \mathbf{h}_{\text{c},k}^{\text{LoS}} + \sqrt{\frac{1}{1+\beta_\text{IC}}} \mathbf{h}_{\text{c},k}^{\text{NLoS}}, \label{eq:54}
\end{equation}
\begin{equation}
\mathbf{g}_\text{e} = \sqrt{\frac{\beta_\text{IE}}{1+\beta_\text{IE}}} \mathbf{g}_\text{e}^{\text{LoS}} + \sqrt{\frac{1}{1+\beta_\text{IE}}} \mathbf{g}_\text{e}^{\text{NLoS}}, \label{eq:55}
\end{equation}
with \(\beta_\text{BI} = 0.5\), \(\beta_\text{IC} = 0.5\), and \(\beta_\text{IE} = 0.5\) are the Rician factors for the BS to IRS, IRS to SCU, and IRS to EVE links, respectively. \(\mathbf{G}_t^{\text{LoS}} \in \mathbb{C}^{N \times M_t}\), \(\mathbf{G}_r^{\text{LoS}} \in \mathbb{C}^{M_r \times N}\), \(\mathbf{h}_{\text{c},k}^{\text{LoS}} \in \mathbb{C}^{N \times 1}\), \(\mathbf{g}_\text{e}^{\text{LoS}} \in \mathbb{C}^{N \times 1}\), and \(\mathbf{G}_t^{\text{NLoS}} \in \mathbb{C}^{N \times M_t}\), \(\mathbf{G}_r^{\text{NLoS}} \in \mathbb{C}^{M_r \times N}\), \(\mathbf{h}_{\text{c},k}^{\text{NLoS}} \in \mathbb{C}^{N \times 1}\), \(\mathbf{g}_\text{e}^{\text{NLoS}} \in \mathbb{C}^{N \times 1}\) are the LoS and \ac{NLoS} (Rayleigh fading with each entry being a CSCG random variable with zero mean and unit variance) components, respectively.
The Rayleigh fading model is employed for the direct links, i.e., BS to SCUs, \(\mathbf{h}_{\text{d},k}\), and BS to EVE links, \(\mathbf{h}_\text{e}\).

\begin{figure}[!t]
  \begin{center}
  \includegraphics[width=3.2in]{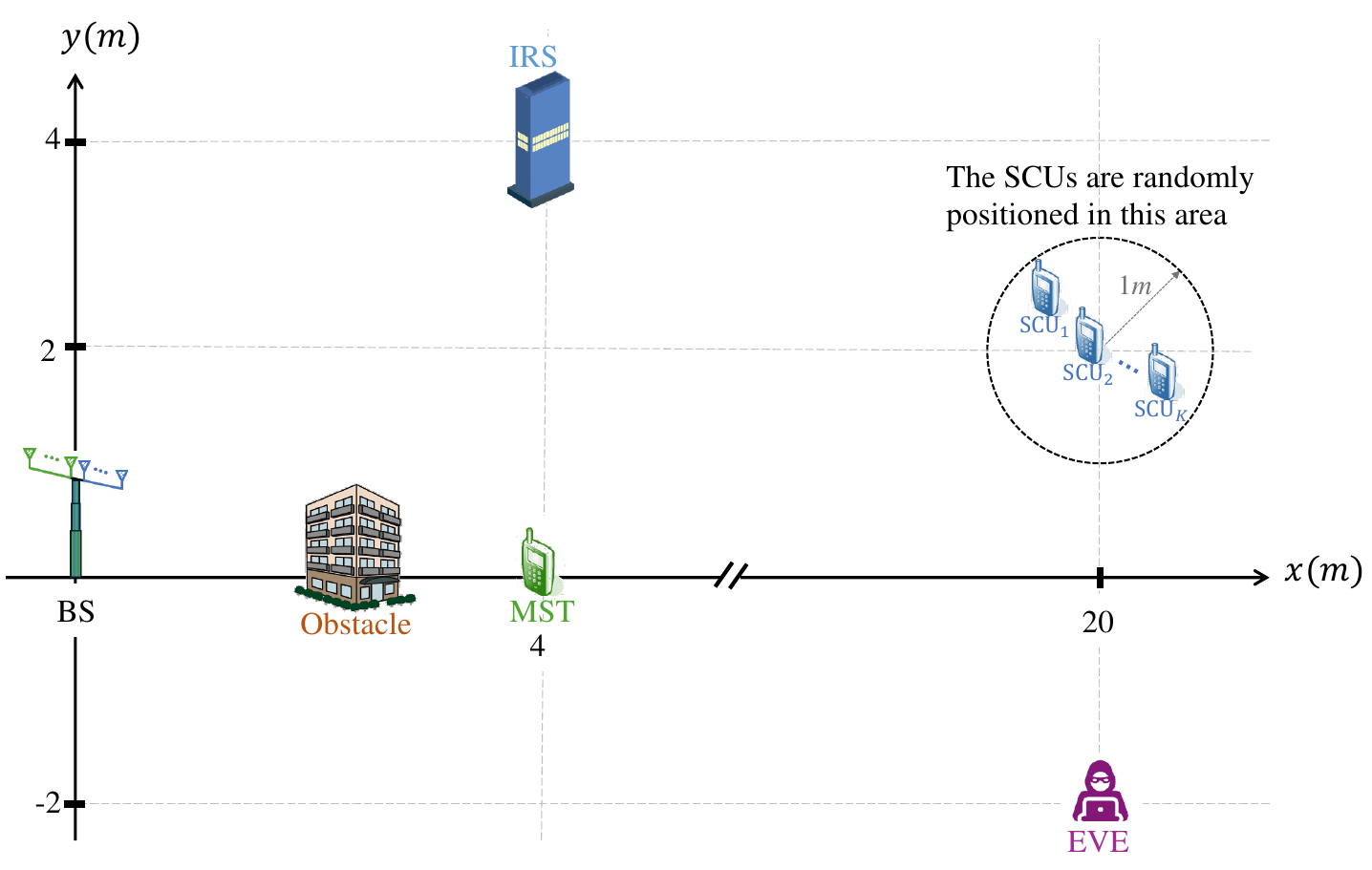}\\
  \caption{The spatial arrangement of the SCUs, BS, EVE, MST, and IRS within the simulation environment.}\label{simulation_setup}
  \end{center}
\end{figure}

We consider a DeepSC-enabled semantic text transmission setup for SC \cite{XIE21}. We set the average number of semantic symbols per data segment $\kappa = 5$, with the corresponding parameters for the generalized logistic function in \eqref{eq_8} defined as $A_{\kappa,1} = 0.37$, $A_{\kappa,2} = 0.98$, $C_{\kappa,1} = 0.25$, and $C_{\kappa,2} = -0.79$ \cite{T15}. 
The stopping thresholds of the algorithms are set as \(\delta_\text{GSS}=\delta_\text{AO} = 10^{-4}\) and the results are averaged over 200 random channel realizations \cite{Hamid23_2}. Other system parameters for BS, IRS, SCUs, MST, and EVE are detailed in Table~\ref{Table_1} and Fig.~\ref{simulation_setup}, unless otherwise specified.

\subsection{Convergence Behaviour of the Proposed Overall Algorithm}
\begin{figure}[!t]
\centering
\includegraphics[width=3.0in]{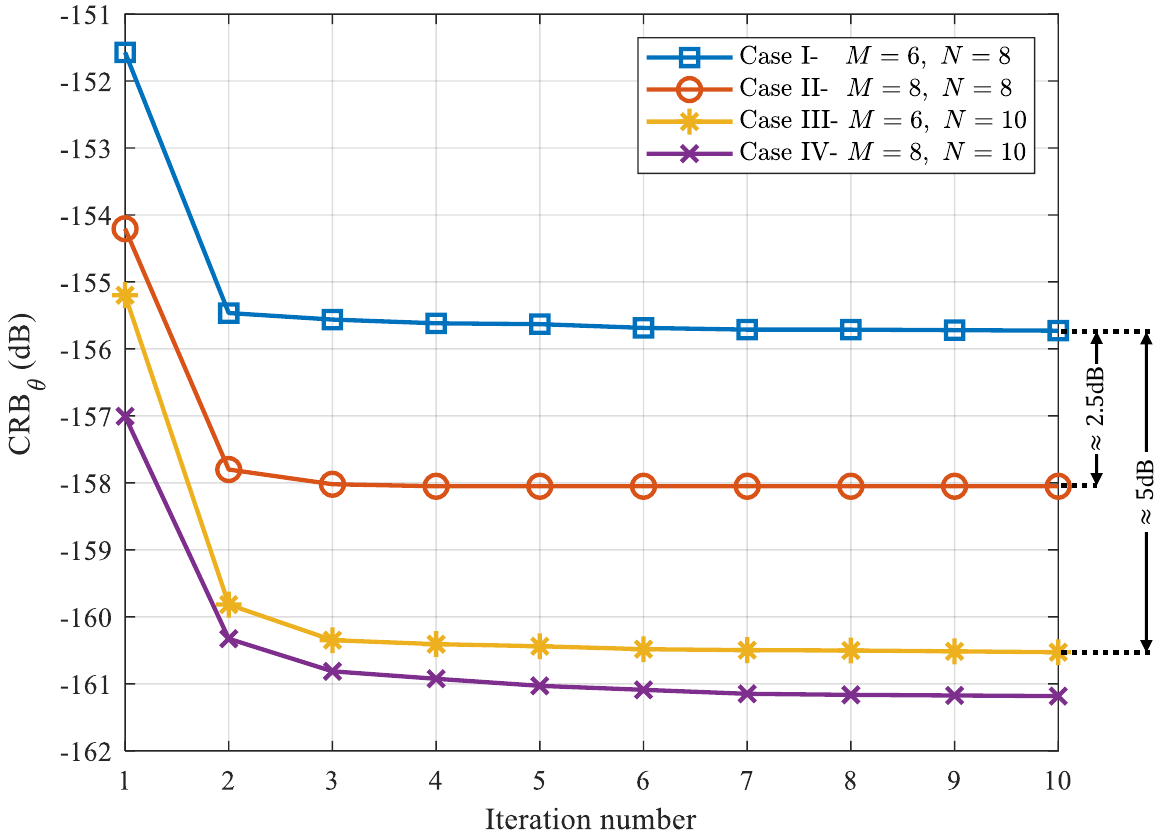}
\vspace{-.28cm}
\caption{Convergence behaviour of the proposed algorithm.}
\label{fig_convergence}
\end{figure}

First, we study the convergence behaviour of the proposed algorithm for problem ($\mathcal{P}.5.2$) as described in Algorithm~\ref{Algorithm2}. Fig.~\ref{fig_convergence} shows the CRB versus the number of iterations for different numbers of transmitter and receiver antennas in the BS, $M_t=M_r=M$, and IRS reflecting elements $N$. Specifically, four cases are considered: Case~I with $M = 6$ and $N = 8$; Case~II with $M = 8$ and $N = 8$; Case~III with $M = 6$ and $N = 10$; Case~IV with $M = 8$ and $N = 10$. 

As can be observed from Fig.~\ref{fig_convergence}, in all four cases, the proposed algorithm converges rapidly to a stationary point within 2-4 iterations, demonstrating the efficiency of our proposed algorithm.
It is noteworthy that the number of iterations required by the proposed algorithm is little affected by changing $M$ and $N$. Specifically, in Case~I, the CRB achieves the saturation value within 2 iterations, while Case~IV requires 5 iterations to converge. This is because of the enlarged solution space with increased $M$ and $N$.
Additionally, for the considered channel realization, comparing Case~I and Case~III, adding two elements to the IRS can decrease the CRB by approximately 5~dB, whereas comparing Case~I and Case~II, adding two transmitter and receiver antennas to the BS results in a smaller reduction in the CRB (approximately 2.5~dB). This is due to the fact that the CRB is primarily influenced by the IRS-MST link when direct links between the BS and the MST are severely blocked by obstacles.

\subsection {Performance Analysis of the Proposed Algorithm}
Next, we compare the performance of our proposed secure IRS-assisted ISASC design (labelled as Proposed approach) based on the method described in Algorithm~\ref{Algorithm1} in Section~\ref{Problem_Formulation} with the following baseline (BL) schemes:
\\\textit{BL~I - Single signal:} The BS employs a single BF vector for communication, sensing, and protection against eavesdropping. The BS BF design in this scheme is similar to our proposed approach as it corresponds to problem~$(\mathcal{SP}1)$ but without DSS and AN, i.e., $\mathbf{w}_\text{s}=\mathbf{0}$, $\mathbf{w}_\text{n}=\mathbf{0}$.
\\\textit{BL~II - MRT-based design:} Optimizes the BS BF vectors for each signal radiated towards the desired direction using the maximum-ratio transmission (MRT) scheme, i.e., $\mathbf{w}_{\text{c},k}^\text{MRT}=\alpha_{\text{c},k}{{\mathbf{h}}_{\text{Bc},k}^\dagger} / {\|{\mathbf{h}}_{\text{Bc},k}\|}$, $\mathbf{w}_{\text{s}}^\text{MRT}=\alpha_{\text{s}}{\hat{\mathbf{h}}_{\text{Bs}}^\dagger}/{\|\hat{\mathbf{h}}_{\text{Bs}}\|}$, and $\mathbf{w}_{\text{n}}^\text{MRT}=\alpha_{\text{n}}{\hat{\mathbf{h}}_{\text{Be}}^\dagger}/{\|\hat{\mathbf{h}}_{\text{Be}}\|}$, where $\bm{\alpha}\triangleq[\alpha_{\text{c},k}, \alpha_\text{s},\alpha_\text{n}]^\top$, $k\in\mathcal{K}$ is the power allocation vector, containing the power allocation variables for SCU, MST, and AN, respectively. Thus, in this scheme, the optimization of the BS BF vectors in ($\mathcal{SP}1$) simplifies to optimizing the power allocation vector, as 
\begin{subequations}
\begin{align}
(\mathcal{SP}1-&\text{BL~II}):\ \underset{\bm{\alpha\in \mathbb{R}^{(K+2)\times1}}}{\Min} \ {\rm CRB}_\theta, \label{eqSP1BLII_a} \\
\text{s.t.} \quad &\mathbf{w}_{\text{c},k}=\alpha_{\text{c},k}\frac{\hat{\mathbf{h}}_{\text{Bc},k}^\dagger} {\|\hat{\mathbf{h}}_{\text{Bc},k}\|},\ k\in\mathcal{K}, \label{eqSP1BLII_b} \\
& \mathbf{w}_{\text{s}}=\alpha_{\text{s}}\frac{\hat{\mathbf{h}}_{\text{Bs}}^\dagger}{\|\hat{\mathbf{h}}_{\text{Bs}}\|}, \ \mathbf{w}_{\text{n}}=\alpha_{\text{n}}\frac{\hat{\mathbf{h}}_{\text{Be}}^\dagger}{\|\hat{\mathbf{h}}_{\text{Be}}\|}, \label{eqSP1BLII_d} \\
& \| \bm{\alpha}\|^2\leq P_\text{max}, \label{eqSP1BLII_e} \\
& \eqref{eqP4_b}, \eqref{eqP4_c}. \nonumber 
\end{align}
\end{subequations}
Therefore, we optimize the IRS phase shift vector, $\mathbf{v}$, in ($\mathcal{SP}2$) alternately with the power allocation vector, $\bm{\alpha}$, in ($\mathcal{SP}1-\text{BL~II}$).
\\\textit{BL~III - 	Isotropic transmission:} In this scheme, the BS employs isotropic transmission by setting the BS BF vectors as $\mathbf{w}_{\text{c},k}^\text{iso}={\alpha_{\text{c},k}\mathbf{I}_{M_t\times1}}/{M_t}$, $\mathbf{w}_\text{s}^\text{iso}={\alpha_\text{s}\mathbf{I}_{M_t\times1}}/{M_t}$, and $\mathbf{w}_\text{n}^\text{iso}={\alpha_\text{n}\mathbf{I}_{M_t\times1}}/{M_t}$. Consequently, this scheme simplifies the optimization of the BS BF vectors in ($\mathcal{SP}1$) to the optimization of the power allocation variables, as follows:
\begin{subequations}
\begin{align}
(\mathcal{SP}1-&\text{BL~III}):\ \underset{\bm{\alpha}}{\Min} \ {\rm CRB}_\theta, \label{eqSP1BLIII_a} \\
\text{s.t.} \quad &\mathbf{w}_{\text{c},k}=\alpha_{\text{c},k}\frac{\mathbf{I}_{M_t\times1}}{M_t},\ k\in\mathcal{K}, \label{eqSP1BLIII_b} \\
& \mathbf{w}_{\text{s}}=\alpha_{\text{s}}\frac{\mathbf{I}_{M_t\times1}}{M_t},\ \mathbf{w}_{\text{n}}=\alpha_{\text{n}}\frac{\mathbf{I}_{M_t\times1}}{M_t}, \label{eqSP1BLIII_d} \\
& \eqref{eqSP1BLII_e}, \eqref{eqP4_b}, \eqref{eqP4_c}. \nonumber 
\end{align}
\end{subequations}
And alternatively optimizes the IRS phase shift vector, $\mathbf{v}$, in ($\mathcal{SP}2$) with power allocation vector $\bm{\alpha}$ in ($\mathcal{SP}1-\text{BL~III}$).
\\\textit{BL~IV - Optimal BS BF only:} we set the phase shift elements in IRS reflection vector, i.e., $\phi_n$, randomly in $[0,2\pi]$. However, the BS BF vectors are optimized by using the optimal results obtained by solving ($\mathcal{SP}1$).

\begin{figure}[!t]
\centering
\includegraphics[width=3.25in]{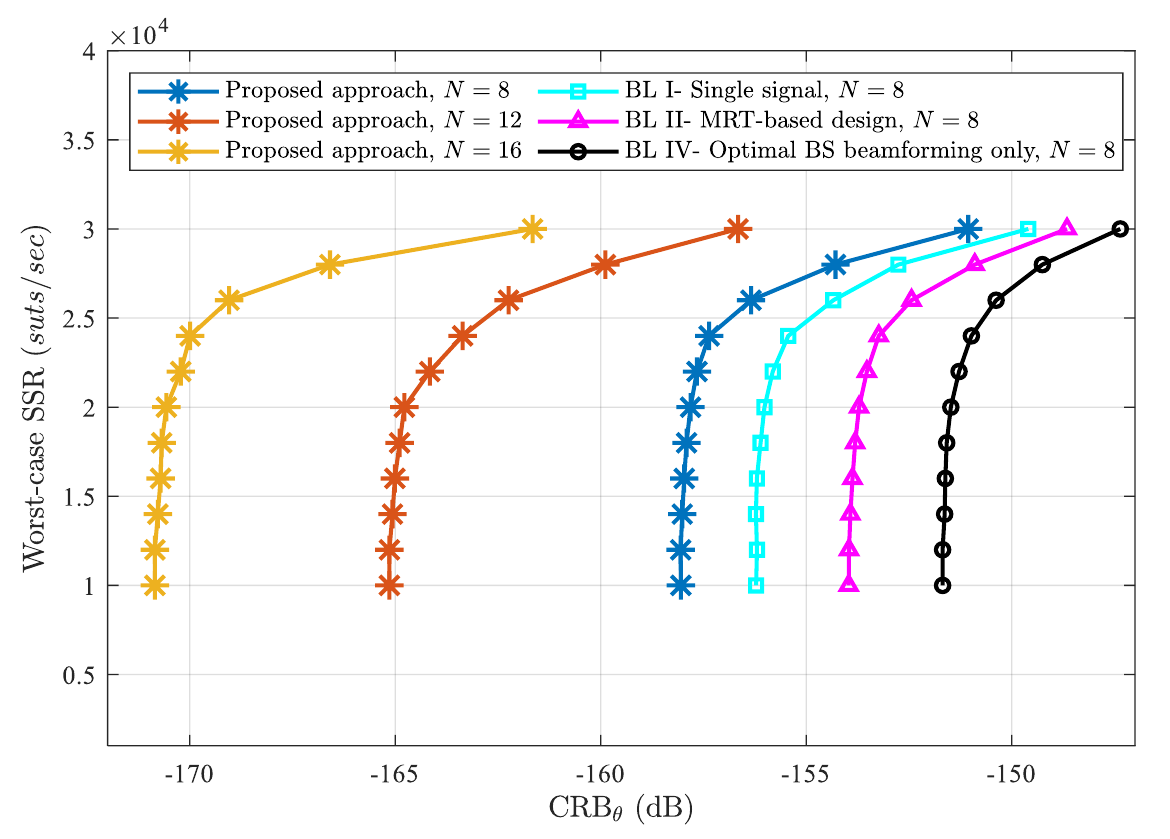}
\caption{Trade-off region between SSR and CRB for the proposed approach with various numbers of IRS reflecting elements, $N$, and different comparison baselines.}
\label{fig4}
\end{figure}

\subsubsection {Trade-off region between SSR and CRB}
Fig.~\ref{fig4} investigates the trade-off region between the worst-case SSR, i.e., $\min_{k \in \mathcal{K}} \operatorname{SSR}_k$, and the CRB for various numbers of IRS reflecting elements, $N$, with baseline comparisons at $N=8$.
Fig.~\ref{fig4} shows that a more stringent SSR requirement results in a higher CRB value for the estimation of the target DoA due to more energy being directed towards SCU, confirming the inherent conflict between minimizing the CRB and maximizing the SSR.
Moreover, it can be seen that increasing the number of IRS reflecting elements, $N$, significantly expands the achievable trade-off region. This demonstrates that deploying an IRS with low-cost reflecting elements is beneficial not only for enhancing sensing accuracy, but also for improving security.
Additionally, the proposed approach outperforms BL~I, where the BS uses a single signal. This indicates the advantages of incorporating the DSS and AN signals, which provide more degrees of freedom (DoF) for the BS to optimize the system performance.
Furthermore, compared to the BL~II, our approach achieves approximately 5~dB lower CRB at an SSR of $2.4\times{10}^4$,  emphasizing the joint optimization of the communication transmit BF vectors, the sensing BF vector, and the AN BF vector are essential to fully reap the gains.
Finally, it is observed that the proposed approach performs better compared to BL~IV, achieving approximately 7~dB lower CRB at an SSR of $2.4\times{10}^4$, due to the optimal IRS phase shift configuration. The reason is that in the proposed approach, the communication signal and DSS beams are optimally directed toward the SCU and MST, respectively, fully exploiting the potential of IRS gain.

\begin{figure}[!t]
\centering
\includegraphics[width=3.5in]{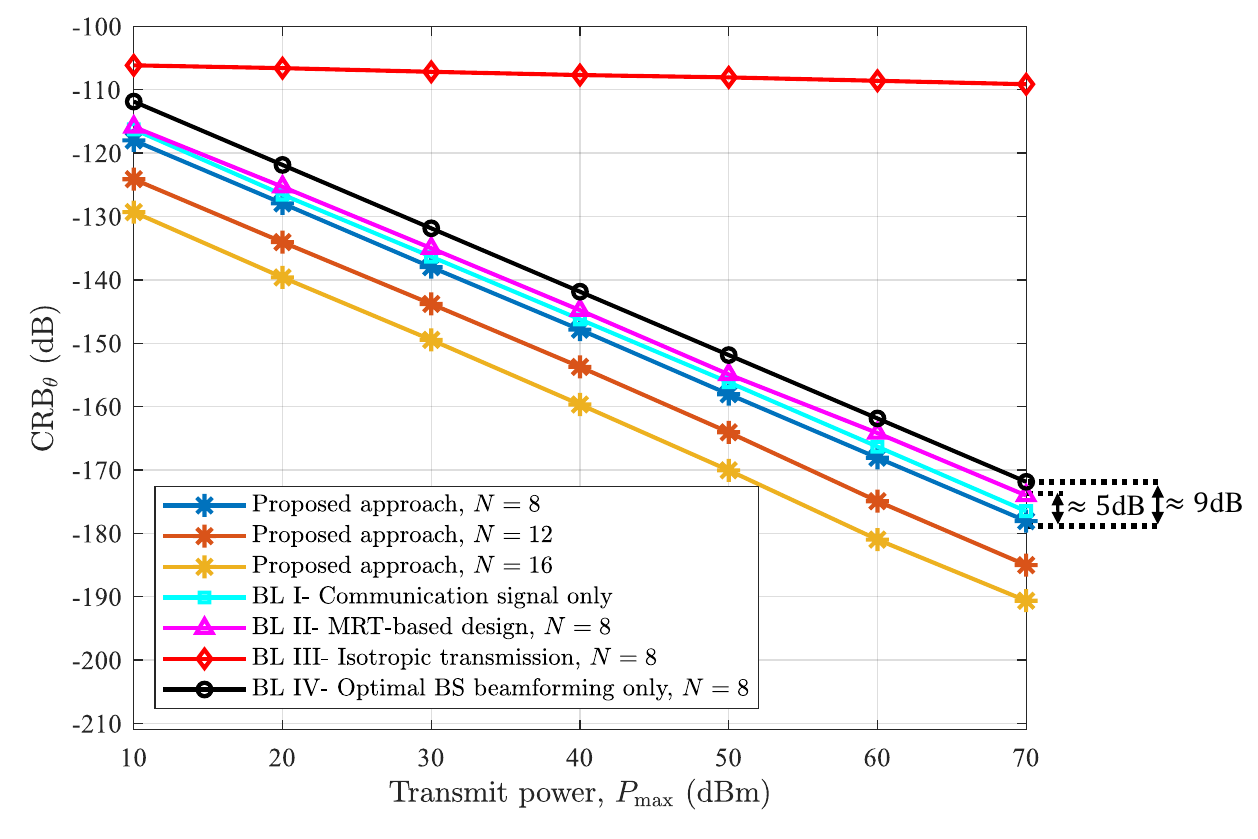}
\caption{CRB versus maximum BS’s transmission power (dBm) at a worst-case SSR of $1 \times 10^4$.}
\label{fig5}
\end{figure}

\subsubsection {CRB Versus Transmit Power}
Next, Fig.~\ref{fig5} compares the ${\rm CRB}_\theta$ using our proposed approach across various numbers of IRS reflecting elements, $N$, and baseline schemes with $N=8$, versus the maximum BS transmission power, $P_\text{max}$, at a worst-case SSR of $1 \times 10^4$.
First, it is observed that the \ac{CRB} (in dB) decreases monotonically in a linear manner w.r.t. $P_\text{max}$ (in dBm), due to its inverse proportionality to $P_\text{max}$ as shown in \eqref{eq_14}.
Additionally, it is noted that our proposed scheme consistently achieves the lowest CRB across the entire transmit power regime compared to other baseline schemes, demonstrating the effectiveness of the proposed IRS-assisted secure ISASC system design for jointly optimizing the BS beamformers and the IRS phase shift.
Furthermore, a significant gap is observed between BL~III, where the BS employs isotropic transmission BF and other schemes. This gap highlights the BF limitations of isotropic strategy compared to other approaches, demonstrating the importance of directional BF in enhancing system performance and efficiency.
Moreover, by comparing the proposed approach and BL~IV at $P_\text{max}=70$~dBm, it is observed that when the IRS phase shift is optimized, performance increases by 9~dB compared to when the IRS phase shift is random. This is due to the fact that in the former case, the DSS benefits from the optimal-phase IRS link, whereas in the latter case, the IRS phases are randomly configured.
Finally, when comparing the proposed approach and BL~II, in which the BS transmit beamformers are separately optimized to maximize the SNR for each direction using MRT, there is a 5~dB improvement in performance. This demonstrates the effectiveness of our proposed joint BF design in minimizing the CRB.

\subsubsection {CRB Versus Number of IRS Reflecting Elements}
\begin{figure}[!t]
\centering
\includegraphics[width=3.2in]{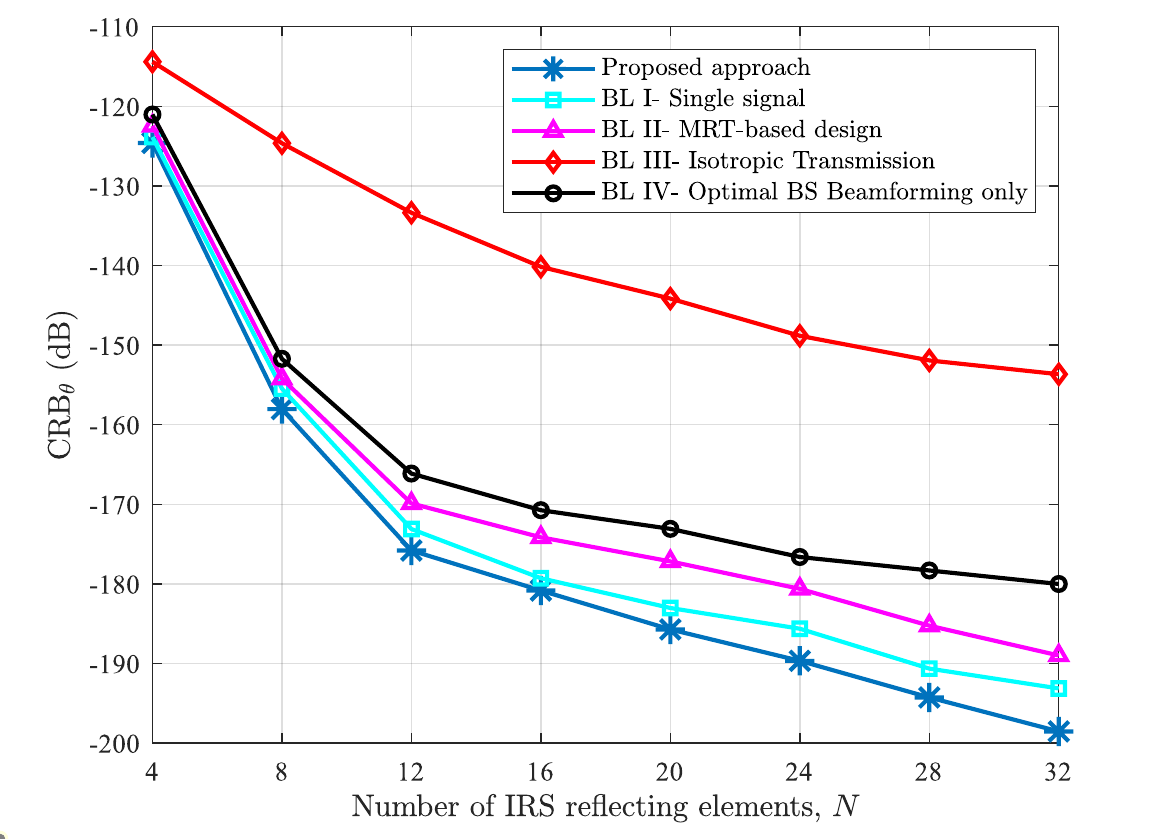}
\caption{CRB (dB) versus the number of IRS reflecting elements, $N$, with $P_\text{max}=50$~dBm  at a worst-case SSR of $1 \times 10^4$.}
\label{fig6}
\end{figure}

We further study the CRB versus the number of reflecting elements at the IRS, $N$, in Fig.~\ref{fig6} for different cases, where $P_\text{max}$ is set to $50$~dBm and the other settings are the same as in Fig. ~\ref{fig5}.
As illustrated in Fig.~\ref{fig6}, the CRB obtained by different approaches decreases almost exponentially with increasing $N$. Specifically, increasing $N$ from 4 to 32 in the proposed approach results in a $74$~dB reduction in CRB values. This demonstrates the substantial reflective BF gains afforded by the IRS.
Additionally, the performance gap between our proposed approach and BL~IV becomes more pronounced with larger $N$. This enhancement is due to the greater DoF available for the IRS to adjust the phase shifts, which contributes to lower CRB values and improved performance when the IRS phase shifts are optimally tuned.
Furthermore, our proposed approach continues to outperform BL~I, which only employs a single signal, further highlighting the benefits of incorporating DSS and AN signals.
Finally, as $N$ increases, the performance gap between our proposed approach and BL~II also widens, which again underscores the advantages of a joint design of BS transmit beamformers.

\subsection {Comparison of Secrecy Performance in SC and Bit-Oriented Communication (BC) Systems}
To gain more insight, the proposed SC scheme is compared to the conventional bit-oriented communication (BC) scheme in terms of secrecy performance.
In BC, both the source coding and the bit transmission processes must be considered for fair comparisons. 
Data in BC are mapped to bits via a source encoder, which can loosely be considered as semantic symbols, albeit with less semantic information than those in SC.
Therefore, the equivalent rate for the BC system can be expressed as \cite[eq.~(11)]{YAN23}:
\begin{equation}\label{eq_57}
\operatorname{BR}^\chi_k(\mathbf{w}, \mathbf{v}) = \frac{BI}{\mu L_\text{s}} \operatorname{R}(\gamma^\chi_k(\mathbf{w}, \mathbf{v})), \chi\in\{\text{com},\text{eve}\}.
\end{equation}
where $\mu$ is a compression factor that reflects the ability of the source coding scheme to compress the data, and $\operatorname{R}(\gamma)$ is the transmission rate, determined by the SNR and its corresponding channel quality indicator (CQI) \cite{CQI_5G}, as specified in Table 7.2.3-1 of 3GPP TS 36.213 \cite{YAN23}.
Thus the secrecy rate for the $k$-th SCU in BC systems can be calculated as: 
\begin{align}\label{eq_58}
\operatorname{BSR}_k(\mathbf{w}, \mathbf{v}) &= [\operatorname{BR}^\text{com}_k(\mathbf{w}, \mathbf{v}) - \operatorname{BR}^\text{eve}_k(\mathbf{w}, \mathbf{v})]^+.
\end{align}

\begin{figure}[!t]
\centering
\includegraphics[width=3.2in]{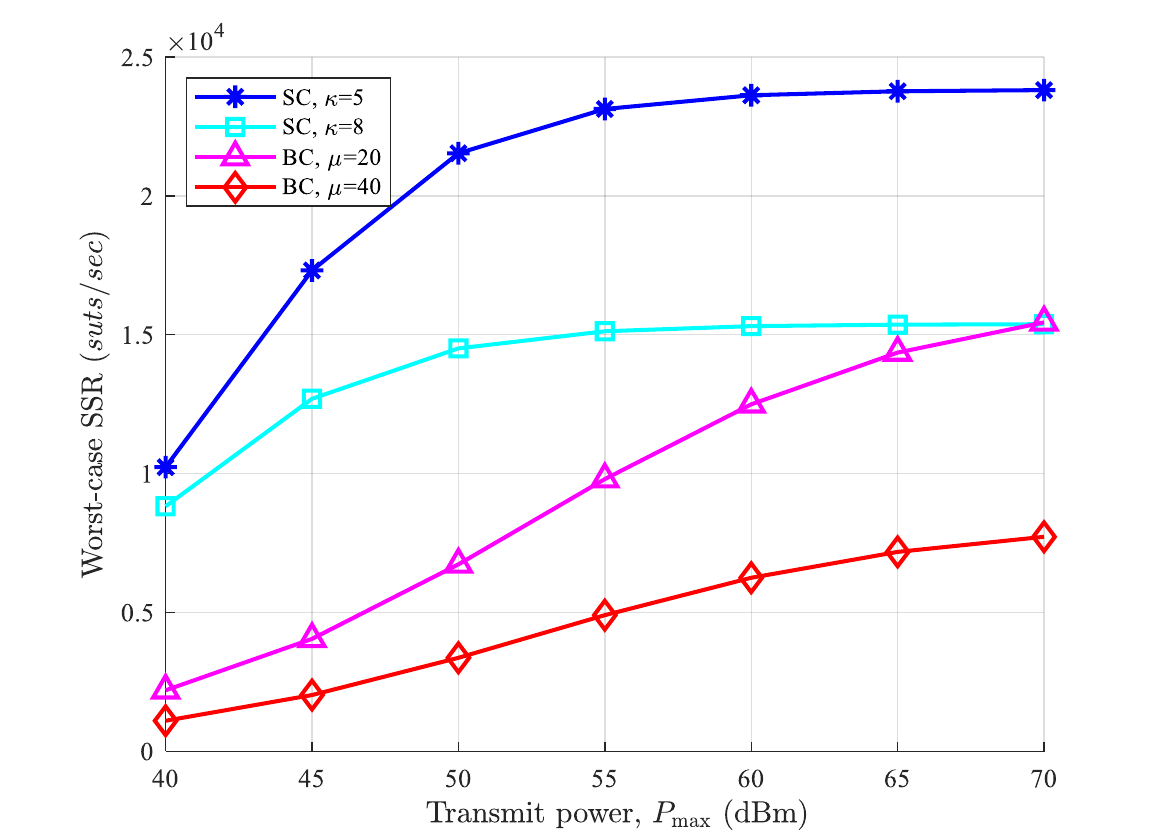}
\caption{Secrecy performance comparison between SC and BC systems, with ${\rm CRB}_\theta = -150$ dB, and $N = 8$.}
\label{fig7}
\end{figure}

Fig.~\ref{fig7} compares the worst-case secrecy rate in SC, i.e., $\operatorname{SSR}_k(\mathbf{w}, \mathbf{v})$ from \eqref{eq_21}, and in BC, i.e., $\operatorname{BSR}_k(\mathbf{w}, \mathbf{v})$ from \eqref{eq_58} for different values of the average number of encoded semantic symbols per data segment in SC, $\kappa$, and the compression factor of the source coding in BC, $\mu$, under ${\rm CRB}_\theta = -150$ dB and $N = 8$ \cite{YAN23}, \cite{Compare}.   
As shown in Fig.~\ref{fig7}, the SSR decreases as $\kappa$ increases, consistent with the discussion of \eqref{eq_21}, which indicates that a higher $\kappa$ leads to a reduced equivalent semantic rate and thereby decreases the semantic secrecy rate.

Furthermore, we observe that as $\kappa$ increases, requiring more semantic symbols to encode each data segment, the performance advantage of SC over BC decreases, especially at $P_{\text{max}} \geq 70$ dBm, with $\kappa = 5$ and $\mu = 20$, where BC outperforms SC.  
This underscores the critical need for efficient designs of semantic encoders and decoders that minimize the required number of semantic symbols while preserving information accuracy.

\section{Conclusion} \label{Conclusion}
In this paper, we formulated an MOOP for the joint BS's communication, sensing, and AN BF vectors and the IRS's phase shift vector design in IRS-assisted ISASC systems to explore the trade-off region between CRB minimization and SSR maximization.
To address this complex and non-convex problem of MOOP, we initially reformulated it into a tractable SOOP using the $\epsilon$-constraint method. 
Subsequently, the problem was divided into two manageable sub-problems, focusing separately on optimizing the BS BF vectors and the IRS phase shift vector. 
To address the non-convex nature of the sub-problems, we applied the SDR technique to transform the non-convex constraints into convex forms and then solved them using SDP. Moreover, we utilized the successive SCA technique to iteratively approximate the non-concave CRB objective function as concave, enabling efficient convergence to a locally optimal solution.
Finally, through iterative solutions of two sub-problems, we accomplished an efficient joint optimization of the system parameters.
Our simulation results demonstrated the efficacy of the overall proposed algorithm, showing that it achieves a better balance between SSR and CRB compared to the baseline schemes.
Furthermore, our performance evaluation highlighted the significant impact of IRS elements in enhancing the system’s ability to control the electromagnetic environment, thereby improving the precision of parameter estimation and overall system performance.


\bibliographystyle{IEEEtran}
\bibliography{IEEEabrv,Bibliography}

\begin{thebibliography}{10}
\providecommand{\url}[1]{#1}
\csname url@rmstyle\endcsname
\providecommand{\newblock}{\relax}
\providecommand{\bibinfo}[2]{#2}
\providecommand\BIBentrySTDinterwordspacing{\spaceskip=0pt\relax}
\providecommand\BIBentryALTinterwordstretchfactor{4}
\providecommand\BIBentryALTinterwordspacing{\spaceskip=\fontdimen2\font plus
\BIBentryALTinterwordstretchfactor\fontdimen3\font minus \fontdimen4\font\relax}
\providecommand\BIBforeignlanguage[2]{{%
\expandafter\ifx\csname l@#1\endcsname\relax
\typeout{** WARNING: IEEEtran.bst: No hyphenation pattern has been}%
\typeout{** loaded for the language `#1'. Using the pattern for}%
\typeout{** the default language instead.}%
\else
\language=\csname l@#1\endcsname
\fi
#2}}
\renewcommand\BIBentryALTinterwordstretchfactor{4}

\bibitem{WCNC2025}
\BIBentryALTinterwordspacing
H.~Amiriara, M.~Mirmohseni, A.~Elzanaty, Y.~Ma, and R.~Tafazolli, ``A physical layer security framework for integrated sensing and semantic communication systems,'' in \emph{IEEE WCNC}, Milan, Italy, 2025, accepted paper. [Online]. Available: \url{https://arxiv.org/abs/2410.06208}
\BIBentrySTDinterwordspacing

\bibitem{Alwis2021}
C.~D. Alwis, A.~Kalla, Q.-V. Pham, P.~Kumar, K.~Dev, W.-J. Hwang, and M.~Liyanage, ``Survey on {6G} frontiers: Trends, applications, requirements, technologies and future research,'' \emph{{IEEE} Open J. Commun. Soc.}, vol.~2, pp. 836--886, 2021.

\bibitem{T1}
H.~Jia, X.~Li, and L.~Ma, ``Physical layer security optimization with {Cramér–Rao} bound metric in {ISAC} systems under sensing-specific imperfect {CSI} model,'' \emph{{IEEE} Trans. Veh. Technol.}, vol.~73, no.~5, pp. 6980--6992, 2024.

\bibitem{T2}
I.~W.~G. da~Silva, D.~P.~M. Osorio, and M.~Juntti, ``Privacy performance of {MIMO} dual-functional radar-communications with internal adversary,'' in \emph{2023 ICC Workshops}, 2023, pp. 1118--1123.

\bibitem{T3}
D.~Xu, X.~Yu, D.~W.~K. Ng, A.~Schmeink, and R.~Schober, ``Robust and secure resource allocation for {ISAC} systems: A novel optimization framework for variable-length snapshots,'' \emph{{IEEE} Trans. Commun.}, vol.~70, no.~12, pp. 8196--8214, 2022.

\bibitem{T4}
Z.~Yang, D.~Li, N.~Zhao, Z.~Wu, Y.~Li, and D.~Niyato, ``Secure precoding optimization for {NOMA}-aided integrated sensing and communication,'' \emph{{IEEE} Trans. Commun.}, vol.~70, no.~12, pp. 8370--8382, 2022.

\bibitem{T5}
N.~Su, F.~Liu, and C.~Masouros, ``Secure radar-communication systems with malicious targets: Integrating radar, communications and jamming functionalities,'' \emph{{IEEE} Trans. Wireless Commun.}, vol.~20, no.~1, pp. 83--95, 2021.

\bibitem{T6}
------, ``Sensing-assisted eavesdropper estimation: An {ISAC} breakthrough in physical layer security,'' \emph{{IEEE} Trans. Wireless Commun.}, vol.~23, no.~4, pp. 3162--3174, 2024.

\bibitem{T7}
H.~Zhao, F.~Wu, W.~Xia, Y.~Zhang, Y.~Ni, and H.~Zhu, ``Joint beamforming design for {RIS}-aided secure integrated sensing and communication systems,'' \emph{{IEEE} Commun. Lett.}, vol.~27, no.~11, pp. 2943--2947, 2023.

\bibitem{T8}
Z.~Xing, R.~Wang, and X.~Yuan, ``Reconfigurable intelligent surface aided physical-layer security enhancement in integrated sensing and communication systems,'' \emph{{IEEE} Trans. Veh. Technol.}, vol.~73, no.~4, pp. 5179--5196, 2024.

\bibitem{T9}
C.~Wang, C.-C. Wang, Z.~Li, D.~W.~K. Ng, K.-K. Wong, N.~Al-Dhahir, and D.~Niyato, ``{STAR}-{RIS}-enabled secure dual-functional radar-communications: Joint waveform and reflective beamforming optimization,'' \emph{{IEEE} Trans. Inf. Forensics Security}, vol.~18, pp. 4577--4592, 2023.

\bibitem{T10}
F.~Xia, Z.~Fei, X.~Wang, P.~Liu, J.~Guo, and Q.~Wu, ``Joint waveform and reflection design for sensing-assisted secure {RIS}-based backscatter communication,'' \emph{{IEEE} Wireless Commun. Lett.}, vol.~13, no.~5, pp. 1523--1527, 2024.

\bibitem{T11}
M.~Hua, Q.~Wu, W.~Chen, O.~A. Dobre, and A.~L. Swindlehurst, ``Secure intelligent reflecting surface-aided integrated sensing and communication,'' \emph{{IEEE} Trans. Wireless Commun.}, vol.~23, no.~1, pp. 575--591, 2024.

\bibitem{T12}
J.~Chu, Z.~Lu, R.~Liu, M.~Li, and Q.~Liu, ``Joint beamforming and reflection design for secure {RIS}-{ISAC} systems,'' \emph{{IEEE} Trans. Veh. Technol.}, vol.~73, no.~3, pp. 4471--4475, 2024.

\bibitem{T13}
A.~A. Salem, M.~H. Ismail, and A.~S. Ibrahim, ``Active reconfigurable intelligent surface-assisted {MISO} integrated sensing and communication systems for secure operation,'' \emph{{IEEE} Trans. Veh. Technol.}, vol.~72, no.~4, pp. 4919--4931, 2023.

\bibitem{T14}
C.~Jiang, C.~Zhang, C.~Huang, J.~Ge, J.~He, and C.~Yuen, ``Secure beamforming design for {RIS}-assisted integrated sensing and communication systems,'' \emph{{IEEE} Wireless Commun. Lett.}, vol.~13, no.~2, pp. 520--524, 2024.

\bibitem{Tnew}
Y.~Xiu, W.~Lyu, P.~L. Yeoh, Y.~Li, Y.~Ai, and N.~Wei, ``Secure enhancement for {RIS}-aided {UAV} with {ISAC}: Robust design and resource allocation,'' \emph{arXiv preprint arXiv:2409.16917}, 2024.

\bibitem{T15}
X.~Mu and Y.~Liu, ``Semantic communication-assisted physical layer security over fading wiretap channels,'' in \emph{ICC 2024}, 2024, pp. 2101--2106.

\bibitem{T16}
Z.~Goldfeld, P.~Cuff, and H.~H. Permuter, ``Semantic-security capacity for the physical layer via information theory,'' in \emph{2016 IEEE Int. Conf. SWSTE}, 2016, pp. 17--27.

\bibitem{T17}
M.~Zhang, Y.~Li, Z.~Zhang, G.~Zhu, and C.~Zhong, ``Wireless image transmission with semantic and security awareness,'' \emph{{IEEE} Wireless Commun. Lett.}, vol.~12, no.~8, pp. 1389--1393, 2023.

\bibitem{T18}
Y.~Wang, W.~Yang, P.~Guan, Y.~Zhao, and Z.~Xiong, ``{TAR}-{RIS}-assisted privacy protection in semantic communication system,'' \emph{{IEEE} Trans. Veh. Technol.}, vol.~73, no.~9, pp. 13\,915--13\,920, 2024.

\bibitem{T22}
Y.~Yang, M.~Shikh-Bahaei, Z.~Yang, C.~Huang, W.~Xu, and Z.~Zhang, ``Joint semantic communication and target sensing for {6G} communication system,'' \emph{arXiv preprint arXiv:2401.17108}, 2024.

\bibitem{XIE21}
H.~Xie, Z.~Qin, G.~Y. Li, and B.-H. Juang, ``Deep learning enabled semantic communication systems,'' \emph{{IEEE} Trans. Signal Process.}, vol.~69, no.~1, pp. 2663--2675, 2021.

\bibitem{GUO15}
L.~Guo, H.~Deng, B.~Himed, T.~Ma, and Z.~Geng, ``Waveform optimization for transmit beamforming with {MIMO} radar antenna arrays,'' \emph{{IEEE} Trans. Antennas Propag.}, vol.~63, no.~2, pp. 543--552, 2015.

\bibitem{SONG23}
X.~Song, J.~Xu, F.~Liu, T.~X. Han, and Y.~C. Eldar, ``Intelligent reflecting surface enabled sensing: {Cramér-Rao} bound optimization,'' \emph{{IEEE} Trans. Signal Process.}, vol.~71, pp. 2011--2026, 2023.

\bibitem{MU23}
X.~Mu, Y.~Liu, L.~Guo, and N.~Al-Dhahir, ``Heterogeneous semantic and bit communications: A semi-{NOMA} scheme,'' \emph{{IEEE} J. Sel. Areas Commun.}, vol.~41, no.~1, pp. 155--169, 2023.

\bibitem{Liu2024}
M.~Zhang, R.~Zhong, X.~Mu, and Y.~Liu, ``Machine learning enabled heterogeneous semantic and bit communication,'' \emph{{IEEE} Trans. Wireless Commun.}, vol.~23, no.~10, pp. 12\,949--12\,963, 2024.

\bibitem{mashhad}
C.~Xu, M.~B. Mashhadi, Y.~Ma, and R.~Tafazolli, ``Semantic-aware power allocation for generative semantic communications with foundation models,'' \emph{arXiv preprint arXiv:2407.03050}, 2024.

\bibitem{YAN23}
L.~Yan, Z.~Qin, R.~Zhang, Y.~Li, and G.~Y. Li, ``Resource allocation for text semantic communications,'' \emph{{IEEE} Wireless Commun. Lett.}, vol.~11, no.~7, pp. 1394--1398, 2022.

\bibitem{channel1}
A.~S. Alwakeel and A.~Elzanaty, ``Semi-blind channel estimation for intelligent reflecting surfaces in massive mimo systems,'' \emph{IEEE Access}, vol.~10, pp. 127\,783--127\,797, 2022.

\bibitem{channel2}
G.~Geraci, M.~Egan, J.~Yuan, A.~Razi, and I.~B. Collings, ``Secrecy sum-rates for multi-user {MIMO} regularized channel inversion precoding,'' \emph{{IEEE} Trans. Commun.}, vol.~60, no.~11, pp. 3472--3482, 2012.

\bibitem{channel3}
A.~Mukherjee and A.~L. Swindlehurst, ``Detecting passive eavesdroppers in the {MIMO} wiretap channel,'' in \emph{IEEE ICASSP}, Kyoto, Japan, 2012, pp. 2809--2812.

\bibitem{NGA05}
P.~Ngatchou, A.~Zarei, and A.~El-Sharkawi, ``Pareto multi objective optimization,'' in \emph{IEEE ISAP}, VA, USA, 2005, pp. 84--91.

\bibitem{CHI13}
K.~Chircop and D.~Zammit-Mangion, ``On $\varepsilon$-constraint based methods for the generation of pareto frontiers,'' \emph{J. Mech. Eng. Autom.}, no.~3, pp. 279--289, 2013.

\bibitem{BER97}
D.~P. Bertsekas, ``Nonlinear programming,'' \emph{J. the Oper. Res. Soc.}, vol.~48, no.~3, pp. 334--334, 1997.

\bibitem{BOYD04}
S.~Boyd, S.~P. Boyd, and L.~Vandenberghe, \emph{Convex Optimization}.\hskip 1em plus 0.5em minus 0.4em\relax Cambridge University Press, 2004.

\bibitem{FUZ05}
F.~Zhang, \emph{The Schur Complement and Its Applications}.\hskip 1em plus 0.5em minus 0.4em\relax Springer-Verlag, 2005.

\bibitem{LUO10}
Z.-q. Luo, W.-k. Ma, A.~M.-c. So, Y.~Ye, and S.~Zhang, ``Semidefinite relaxation of quadratic optimization problems,'' \emph{{IEEE} Signal Process. Mag.}, vol.~27, no.~3, pp. 20--34, 2010.

\bibitem{LI23}
J.~Li, G.~Zhou, T.~Gong, and N.~Liu, ``Beamforming design for active {IRS}-aided {MIMO} integrated sensing and communication systems,'' \emph{{IEEE} Commun. Lett.}, vol.~12, no.~10, pp. 1786--1790, 2023.

\bibitem{HONG16}
M.~Hong, M.~Razaviyayn, Z.-Q. Luo, and J.-S. Pang, ``A unified algorithmic framework for block-structured optimization involving big data: With applications in machine learning and signal processing,'' \emph{{IEEE} Signal Process. Mag.}, vol.~33, no.~1, pp. 57--77, 2016.

\bibitem{complexity1}
S.~Arora and B.~Barak, \emph{Computational complexity: a modern approach}.\hskip 1em plus 0.5em minus 0.4em\relax Cambridge University Press, 2009.

\bibitem{complexity2}
I.~P{\'o}lik and T.~Terlaky, ``Interior point methods for nonlinear optimization,'' \emph{Nonlinear Optimization}, pp. 215--276, 2010.

\bibitem{Hamid_Nomadic}
H.~Amiriara, M.~R. Zahabi, and V.~Meghdadi, ``Power-location optimization for cooperative nomadic relay systems using machine learning approach,'' \emph{{IEEE} Access}, vol.~9, pp. 74\,246--74\,257, 2021.

\bibitem{LI24}
X.~Li, H.~Wang, Y.~Chen, and S.~Sheng, ``Joint resource allocation and reflecting precoding design for {RIS}-assisted {ISAC} systems,'' \emph{{IEEE} Wireless Commun. Lett.}, vol.~13, no.~4, pp. 1193--1197, 2024.

\bibitem{Hamid23_2}
H.~Amiriara, F.~Ashtiani, M.~Mirmohseni, and M.~Nasiri-Kenari, ``{IRS}-user association in {IRS}-aided {MISO} wireless networks: Convex optimization and machine learning approaches,'' \emph{{IEEE} Trans. Veh. Technol.}, vol.~72, no.~11, pp. 14\,305--14\,316, 2023.

\bibitem{CQI_5G}
E.~Chu, J.~Yoon, and B.~C. Jung, ``A novel link-to-system mapping technique based on machine learning for 5{G}/{IoT} wireless networks,'' \emph{Sensors}, vol.~19, no.~5, 2019.

\bibitem{Compare}
P.~Xie, F.~Li, M.~Zhang, W.~Quan, J.~Zhu, and N.~Cheng, ``{STAR}-{RIS} assisted information transmission based on fairness in semantic communication systems,'' \emph{{IEEE} Trans. Wireless Commun.}, vol.~23, no.~11, pp. 17\,007--17\,020, 2024.

\end{thebibliography}

\vfill

\end{document}